\documentclass[twocolumn]{autart}
\usepackage{algorithmic}

\usepackage[square,sort,comma,numbers]{natbib}            
\usepackage{graphicx}          

\usepackage[cmex10]{amsmath}   
\usepackage{array}
\usepackage{latexsym}
\usepackage{amssymb,amsbsy}
\usepackage{multirow}
\usepackage{xcolor}
\usepackage{enumitem}
\usepackage{algorithm}
\usepackage{url}
\usepackage{soul}

\definecolor{myOrange}{rgb}{1,0.5,0}

\newcommand*{\red}{\textcolor{red}}

\newcommand*{\rank}{\mathrm{rank}}
\newcommand{\eio}{e^{i\omega}}
\newcommand{\eiom}{e^{-i\omega}}
\newtheorem{proposition}{Proposition}
\newtheorem{corollary}{Corollary}

\newtheorem{remark}{Remark}
\newtheorem{lemma}{Lemma}
\newtheorem{assumption}{Assumption}

\newtheorem{example}{Example}

\newtheorem{theorem}{Theorem}

\begin{document}

\begin{frontmatter}

\title{Generic identifiability of subnetworks in a linear dynamic network: the full measurement case }

\thanks[footnoteinfo]{\copyright $\langle 2021 \rangle$. This manuscript version is made available under the CC-BY-NC-ND 4.0 \\ license http://creativecommons.org/licenses/by-nc-nd/4.0/}

\thanks[footnoteinfo1]{ Submitted for publication to Automatica, 4 August 2020. Revised 27 July 2021 and 20 October 2021. Accepted for publication 25 October 2021. This project has received funding from the European Research Council (ERC), Advanced Research Grant SYSDYNET, under the European Union’s Horizon 2020 research and innovation programme (grant agreement No 694504).}

\author[add1]{Shengling Shi}, \ead{s.shi@tue.nl}
\author[add1,add2]{Xiaodong Cheng} \ead{xc336@cam.ac.uk} and
\author[add1]{Paul M. J. Van den Hof} \ead{p.m.j.vandenhof@tue.nl}
\address[add1]{Control Systems Group, Department of Electrical Engineering,  Eindhoven University of Technology, \\5600 MB Eindhoven, The Netherlands}
\address[add2]{Department of Engineering, University of Cambridge, Trumpington Street, Cambridge, CB2 1PZ, United Kingdom}

\begin{abstract}
Identifiability conditions for single or multiple modules in a dynamic network specify under which conditions the considered modules can be uniquely recovered from the second-order statistical properties of the measured signals. Conditions for generic identifiability of multiple modules, i.e. a subnetwork, are developed for the situation that all node signals are measured and excitation of the network is provided by both measured excitation signals and unmeasured disturbance inputs. Additionally, the network model set is allowed to contain non-parametrized modules that are fixed, and e.g. reflect modules of which the dynamics are known to the user. The conditions take the form of path-based conditions on the graph of the network model set. Based on these conditions, synthesis results are formulated for allocating external excitation signals to achieve generic identifiability of particular subnetworks. If there are a sufficient number of measured external excitation signals, the formulated results give rise to a generalized indirect type of identification algorithm that requires only the measurement of a subset of the node signals in the network.
\end{abstract}


\end{frontmatter}

\section{Introduction}
Due to the increasing complexity of current technological systems, the study of large-scale interconnected dynamic systems receives considerable attention \cite{mesbahi2010graph,ishizaki2018graph,Cheng2021Survey}.
The resulting dynamic networks can adequately describe a wide class of complex engineering systems appearing in various applications, including multi-robot coordination \citep{mesbahi2010graph}, power grids \citep{ishizaki2018graph} and gene networks \citep{adebayo2012dynamical}.
For data-driven modeling problems in structured dynamic networks, different types of network models have been used. Connecting to prediction-error identification methods, the most popular modeling framework is based on a network of transfer functions, initally introduced in \citep{Goncalves&Warnick:08,van2013identification}, where vertices represent the internal signals, that are available for measurement, and directed edges denote transfer functions which are called modules. While there are alternatives, e.g. in the form of state-space models \citep{haber2014subspace,Yu2018Subspace2}, in this paper we will adhere to the former so-called {\it module} framework.\\
Estimating network properties from data can be addressed in terms of different problem settings. One can e.g. consider the estimation of the network topology, i.e. the interconnection structure of the network \citep{materassi2012problem,sanandaji2011acc,chiuso2012bayesian, hayden2016sparse,ShiTopology,Zorzi&Chiuso:17}. Another problem is to identify a single module in a network with given topology. This includes the selection of internal signals that need to be measured and excited for achieving consistent module estimates \citep{van2013identification,dankers2015,gevers2015identification,Linder&Enqvist_ijc:17,Everitt&Bottegal&Hjalmarsson_Autom:18,gevers2018practical,materassi2019signal,ramaswamy2019local}. Identification of the full network dynamics, for given network topology, is addressed in e.g., \cite{chiuso2012bayesian,weerts2018prediction,Fonken&etal_IFAC:20}.

In this paper, we focus on {\it network identifiability}, which is a concept that is independent of the particular identification method chosen. Based on the results for deterministic network reconstruction problems in \cite{Goncalves&Warnick:08,adebayo2012dynamical}, the concept of global network identifiability was introduced for an identification setting in \cite{Weerts&etal_IFAC:15,HarmAutomatica}, as a property that reflects the ability to distinguish between network models in a parametrized model set based on measured data. In the literature, there are two notions of network identifiability, namely, \textit{global identifiability}  \cite{Weerts&etal_IFAC:15,HarmAutomatica,van2018necessary} that requires \textit{all} model to be distinguishable from the other models in the model set\footnote{There are actually two versions of global identifiability, reflecting whether either one particular model in the set can be distinguished or all models in the set \cite{HarmAutomatica}.}, and \textit{generic identifiability} \cite{bazanella2017identifiability,hendrickx2018identifiability,HarmGraphCDC}, which means that \textit{almost all} models can be distinguished from all other models in the model set. It has been shown in \cite{bazanella2017identifiability,hendrickx2018identifiability} that by considering generic identifiability, the algebraic conditions for identifiability can be recast into path-based conditions on the topology/graph of the network models, which largely simplifies the analysis.

Network identifiability is typically dependent on several structural properties of the model set, such as the network topology, the
modeled correlation structure of process noises, the presence and location of external excitation signals and the availability of measured vertex signals. Conditions for network identifiability have been analyzed for different problem settings. In the setting with full excitation \cite{bazanella2017identifiability,hendrickx2018identifiability,van2018necessary}, all vertices are excited by external excitation signals while only a subset of vertices is measured. In contrast, the full measurement setting in \cite{HarmAutomatica,HarmGraphCDC} assumes that all vertex signals are measured, while only a subset of them is excited. Recent contributions in \cite{bazanella2019network,Shi&Cheng&VandenHof:20,cheng2021necessary} also address the combined situation.

In this paper our objective is to derive path-based conditions for generic identifiability of only a subset of modules (subnetwork) in the network, while we assume all internal signals in the network to be available for measurement, the so-called full measurement case. Additionally, we will require the conditions to be suitable for solving the synthesis problem too, i.e. the allocation of a minimum set of external excitation signals, so as to achieve generic identifiability of the subnetwork.

For analysing this problem, we start from the path-based results for generic identifiability of a subnetwork as presented in \cite{bazanella2017identifiability,hendrickx2018identifiability}. In these works identifiability is defined as a property of a single network, while disturbance signals are not exploited. In our approach we follow up on the more general model-set type of definition of identifiability as introduced in \citep{HarmAutomatica,HarmGraphCDC}, that is more aligned to the use of this concept in an identification setting. Moreover this step allows to incorporate the following attractive features that are all addressed in the current paper: (a) it allows to include the effect of unmeasured disturbance signals in the network as an information source that can contribute to satisfying identifiability conditions; (b) it allows to include modules in the network that are a priori known to the user and thus do not need to be identified. In this setting we will develop novel analysis tools for generic identifiability of subnetworks that are formulated in terms of disconnecting sets in the graph of the network models, and we will show that this leads to a new and effective synthesis procedure for allocating external excitation signals for achieving identifiability of a subnetwork. This synthesis problem was not addressed in  \cite{bazanella2017identifiability,hendrickx2018identifiability,HarmAutomatica,HarmGraphCDC}. As a side result, a new generalized indirect identification method is described that follows immediately form the identifiability conditions, in the situation that a sufficient number of measured excitation signals is present. This method points to a subset of internal signals that in the considered situation would be sufficient for uniquely identifying the subnetwork.

The synthesis problem as formulated above is important for actually designing experimental setups for subnetwork identification. A related synthesis problem for full network identifiability has been addressed in \citep{2019arXivChen}, but requires completely different tools for analysis.

The paper proceeds as follows. After introducing preliminaries and the problem in Section \ref{sec:model}, algebraic and path-based conditions for generic identifiability are formulated in Sections \ref{sec:Algebra} and \ref{sec:AllocateSingleNew}. Disconnecting-set-based conditions are then derived in Section \ref{sec:disconset}, leading to synthesis approaches in Section \ref{sec:synthesis}. Then a generalized indirect identification method that directly follows from the identifiability conditions is presented in Section \ref{sec:indirect}.

Preliminary results of this paper were presented in \cite{ShiReport}. In the current paper comprehensive algebraic and path-based conditions are formalized for a generalized situation, including the step from single modules to subnetworks. Additionally a novel indirect identification method is presented.
\section{Preliminaries and problem formulation} \label{sec:model}
\subsection{Dynamic networks}
The dynamic network models the relation among a set of measured scalar \textit{internal signals} $ \mathcal{W} \triangleq \{w_1(t),\cdots,w_L(t)\}$ with $L = |\mathcal{W}|$ denoting the cardinality of $\mathcal{W}$, a set of measured excitation signals $ \mathcal{R} \triangleq \{r_1(t),\cdots,r_K(t)\}$ with $K = |\mathcal{R}|$, and a set of unmeasured disturbance signals $\{v_1(t),\cdots,v_L(t)\}$. The model is written as
\begin{equation}
w(t) = G(q)w(t)+R(q) r(t)+v(t), \label{eq:model}
\end{equation}
where $G(q)$ and $R(q)$ are matrices of rational transfer operators with delay operator $q^{-1}$, i.e. $q^{-1} w_i(t) = w_i(t-1)$; $w(t)$, $r(t)$ and $v(t)$ are the column vectors that collect all the internal signals, excitation signals and disturbances, respectively. In addition, $v(t)$ is a vector of zero-mean stationary stochastic processes with power spectrum $\Phi_v(\omega)$, which is modeled as a filtered white noise vector $e(t)$ according to:
\begin{equation}
v(t) = H(q)e(t), \label{eq:model1}
\end{equation}
where $e(t)$ has a covariance matrix $\Lambda$. Depending on whether $\Phi_v(\omega)$ is of full rank or not, $H(q)$ can either be square or have more rows than columns \citep{HarmAutomatica}.
Combining \eqref{eq:model} and \eqref{eq:model1} leads to a dynamic network model that is supposed to satisfy the following Assumption \ref{ass1}.
\begin{assumption}\mbox{ }
\label{ass1}
\begin{enumerate}[label=(\alph*)]
\item $G(q)$ has zero diagonal elements, and $G(q)$ is proper and stable;

\item The network is well-posed in the sense that all principal minors of $\lim_{z \to \infty}(I-G(z))$ are non-zero;

\item $[I-G(q)]^{-1}$ is stable;

\item $R(q)$ is a proper and stable rational transfer matrix;

\item $H(q)$ is minimum phase and monic if square; for the non-square case, i.e. when $\Phi_v(\omega)$ is singular, $H(q)$ is structured as $H(q) = \begin{bmatrix} H_a(q) \\ H_b(q) \end{bmatrix}$, with $H_a(q)$ square, monic and minimum phase \cite{HarmAutomatica};

\item The covariance matrix $\Lambda$ of $e(t)$ is positive definite.
\end{enumerate}
\end{assumption}

In the above assumptions, (b) and (c) ensure that every transfer function from external signals to internal signals is proper and stable
\citep{dankers2014system}; the stability of $G(q)$ in (a) guarantees that the noise filter defined later in \eqref{eq:EtoImap} is also inversely stable, which is typical for the modeling of stationary stochastic processes.

Both the excitation signals in $\mathcal{R}$ and the entries (white noises) in $e(t)$ are called \textit{external signals}, and the set of all external signals is denoted by  $\mathcal{X}$. The entries in $G(q)$ are referred to as \textit{modules}. In addition, the dynamic network model leads to mappings from the excitation signals to internal signals:
\begin{equation}
w(t)  =  T_{\mathcal{W} \mathcal{X}}(q) \begin{bmatrix} r(t) \\ e(t) \end{bmatrix}
      =  T_{\mathcal{W} \mathcal{R}}(q) r(t) + \bar{v}(t), \label{eq:EtoImap}
\end{equation}
where $T_{\mathcal{W} \mathcal{X}}(q) \triangleq [I-G(q)]^{-1}X(q)$, $X(q) \triangleq [R(q) \ \ H(q)]$,
$T_{\mathcal{W} \mathcal{R}}(q) \triangleq [I-G(q)]^{-1}R(q)$, $ \bar{v}(t) \triangleq [I - G(q)]^{-1}H(q)e(t)$.\\ The power spectrum $\Phi_{\bar v}(\omega)$ of $\bar{v}(t)$ satisfies $\Phi_{\bar v}(\omega) =[I-G(\eio)]^{-1}H(\eio)\Lambda H^T(\eiom)[I-G(\eiom)]^{-T}.$
When applying common statistical identification methods to \eqref{eq:EtoImap} that are based on first and second moment information of the measured signals, typically the objects $T_{\mathcal{W} \mathcal{R}}(q)$ and $\Phi_{\bar{v}}(\omega)$ can be consistently estimated from measured signals $w(t)$ and $r(t)$, if $r(t)$ is persistently exciting \citep{ljung1987system}. This motivates to use $T_{\mathcal{W} \mathcal{R}}(q)$ and $\Phi_{\bar{v}}(\omega)$ as bases for network identifiability \cite{HarmAutomatica}.

Given subsets $\bar{\mathcal{W}} \subseteq \mathcal{W}$ and $\bar{\mathcal{X}} \subseteq \mathcal{X}$, $T_{\bar{\mathcal{W}}  \bar{\mathcal{X}}}(q)$ denotes a submatrix of $T_{\mathcal{W} \mathcal{X}}(q)$ with the rows and columns corresponding to the signals in $\bar{\mathcal{W}}$ and $\bar{\mathcal{X}}$. If $\bar{\mathcal{W}}$ contains only one signal $w_k$, $T_{\bar{\mathcal{W}}  \bar{\mathcal{X}}}(q)$ is often simply written as $T_{k  \bar{\mathcal{X}}}(q)$. The above notation applies similarly to submatrices of other matrices and vectors.

%
%
\subsection{Model sets}
Network identifiability will be defined on the basis of a network model set, which is introduced as follows.
The network model \eqref{eq:model}, \eqref{eq:model1} is completely specified by a quadruple $M \triangleq (G(q),R(q),H(q),\Lambda)$. By parametrizing the entries of the network matrices in a rational form and then collecting the parameters into a parameter vector $\theta$, a parametrized model set can be defined.

\begin{defn}
Consider a rational parametrization of a network model according to
$$M(\theta) = (G(q,\theta),R(q,\theta),H(q,\theta),\Lambda(\theta)).$$ Then a network model set $\mathcal{M}$ is defined as
$$
\mathcal{M}= \{M(\theta)| \theta \in \Theta \subseteq \mathbb{R}^n  \},
$$
where $M(\theta)$ satisfies Assumption \ref{ass1} for all $\theta \in \Theta$.
\end{defn}
%

In addition, there can be certain entries in $G(q,\theta)$, $R(q,\theta)$ and $H(q,\theta)$ that are fixed/known and thus do not depend on the parameters. These entries are said to be \textit{known} or {\it fixed}, and all models in $\mathcal{M}$ contain the same known entries. The entries in the network matrices that depend on the parameters are said to be \textit{unknown} or {\it parametrized}. For example, the absence of an interconnection between internal signals is represented by a fixed $0$ in $G(q,\theta)$; entries in $G(q,\theta)$ may be particularly designed controllers that are fixed and known. Similarly,
entries in $H(q,\theta)$, $R(q,\theta)$ and $\Lambda(\theta)$ can be fixed, e.g., equal to $1$ or $0$. In the situation of $R(q)$, this implies that it can be specified upfront on which nodes the $r$-signals enter the network.
Note that the dependency of transfer matrices on $q$, $\theta$ and the dependency of power spectra on $\omega$ are sometimes omitted for simplicity of notation.

The structural information of a model set is reflected by a directed graph $\mathcal{G} = (\mathcal{V},\mathcal{E})$, where $\mathcal{V} \triangleq \mathcal{W} \cup  \mathcal{X}$ is a set of vertices representing both the internal signals and the external signals, and $\mathcal{E} \subseteq \mathcal{V} \times \mathcal{V}$ denotes a set of directed edges representing those entries in $G(q,\theta)$ and $X(q,\theta)$ that are not fixed to zero: a directed edge from $w_i$ to $w_j$ exists, i.e. $(w_i,w_j) \in \mathcal{E}$,  if $G_{ji}$ is not fixed to zero; Similarly, $(e_k,w_j) \in \mathcal{E}$ and $(r_p,w_j) \in \mathcal{E}$ if $H_{jk}$ and $R_{jp}$ are not fixed to zero, respectively. In this way, any parametrized model set or network model induces a directed graph $\mathcal{G}$. Note that $w_i$ now represents both a signal and a vertex, and its dependency on $t$ is sometimes omitted for simplicity of notation.\\
For deriving the results in the sequel of this paper, we will need an additional concept.

\begin{defn} \label{def:strucModelSet}
Given a model set $\mathcal{M}$ with its graph $\mathcal{G}$, the set $\mathbb{G}^\star$ is defined as the set of all $G(q)$ matrices that satisfy the following conditions: (i) they meet Assumption \ref{ass1}; (ii) they have the same fixed entries as $G(q,\theta)$ in $\mathcal{M}$; (iii) entries are strictly proper if the corresponding entries in $G(q,\theta)$ are (parametrized to be) strictly proper.
\end{defn}

In Definition~\ref{def:strucModelSet}, $\mathbb{G}^\star$ is the largest possible set of $G(q)$ matrices that have the same fixed entries and feedthrough structure as $G(q,\theta)$ in $\mathcal{M}$. So while the modules $G$ in $\mathcal{M}$ can be restricted in order, the order of entries in $\mathbb{G}^\star$ is not bounded. This implies $\{G(q,\theta)| \theta \in \Theta \} \subseteq \mathbb{G}^\star$. In Assumption~\ref{ass:openset} we will introduce a technical condition on $\{G(q,\theta)| \theta \in \Theta \}$ with respect to $\mathbb{G}^\star$ for proving necessary conditions for identifiability.

%
%
%

\subsection{Network identifiability}
We focus on the identifiability of a subset of unknown modules in one row of the $G$ matrix, i.e. a subnetwork of modules that share the same output node $w_j$. For this output node $w_j$, we define an important set of signals:
\begin{itemize}
    \item $\mathcal{W}_j$: all the internal signals that have unknown directed edges (modules) to $w_j$.
\end{itemize}
For any subset $\bar{\mathcal{W}}_j \subseteq \mathcal{W}_j$, $G_{j\bar{\mathcal{W}}_j}$ denotes a row vector containing a subset of unknown modules in the $j$th row of $G$. For the identifiability of $G_{j\bar{\mathcal{W}}_j}$, we follow the concept of global network identifiability in \citep{HarmAutomatica}, and extend it with a generic version introduced in \cite{bazanella2017identifiability, hendrickx2018identifiability}. In this respect we follow an approach that was suggested in \cite{HarmGraphCDC}.

\begin{defn}
\label{defif} \label{def:defnOriIden}
Given a network model set $\mathcal{M}$, consider any subset $\bar{\mathcal{W}}_j \subseteq \mathcal{W}_j$, a parameter vector $\theta_0 \in \Theta$ and the following implication:
\begin{equation} \label{equivTP}
		\left. \begin{array}{c} T_{\mathcal{W}\mathcal{R}}(q,\theta_0)\!=\! T_{\mathcal{W} \mathcal{R}}(q,\theta_1) \\ \Phi_{\bar v}(\omega,\theta_0)\!=\! \Phi_{\bar v}(\omega,\theta_1) \end{array} \right\}
		\!\Rightarrow
G_{j\bar{\mathcal{W}}_j}(q,\theta_0)\! =\! G_{j\bar{\mathcal{W}}_j}(q,\theta_1)
\end{equation}
for all $\theta_1 \in \Theta$. Then $G_{j\bar{\mathcal{W}}_j}(q,\theta)$ is said to be
\begin{itemize}
\item globally identifiable in $\mathcal{M}$ from $(w,r)$ if the implication (\ref{equivTP}) holds for all $\theta_0 \in \Theta$;
\item generically identifiable in $\mathcal{M}$ from $(w,r)$ if the implication (\ref{equivTP}) holds for almost all $\theta_0 \in \Theta$.
\end{itemize}
\end{defn}
In Definition~\ref{def:defnOriIden}, the notion ``almost all'' excludes a subset of Lebesgue measure zero from $\Theta$. The identifiability concept in Definition~\ref{def:defnOriIden} concerns whether $G_{j\bar{\mathcal{W}}_j}$ is unique given the objects that can typically be identified from the first and second moment information of the measured signals, as motivated following (\ref{eq:EtoImap}).
%
If a model set is not identifiable, any identification method that relies on
the first and second moments for estimating the network can not be guaranteed to (asymptotically) identify a unique network model. In addition, recall that the network matrices in $\mathcal{M}$ may contain known entries, and this restriction of $\mathcal{M}$ may simplify the conditions under which the implication \eqref{equivTP} holds.

To further simplify \eqref{equivTP}, we use the following assumption on the feedthrough terms in $G$.

\begin{assumption}[\citep{HarmAutomatica}]
\label{ass:feedthrough}
 \begin{enumerate}[label=(\alph*)]
 \item either all modules $G(q,\theta)$ are parametrized to be strictly proper, or
 \item the parametrized network model does not contain any algebraic loops\footnote{There exists an algebraic loop around node $w_{n_1}$ if there exists a sequence of integers $n_1,... n_k$ such that $G^{\infty}_{n_1n_2}G^{\infty}_{n_2n_3}... G^{\infty}_{n_kn_1} \neq 0$, with $G^{\infty}_{n_1n_2}:=\lim_{z\rightarrow\infty} G_{n_1n_2}(z)$.}, and  $H^{\infty}(\theta)\Lambda(\theta)[H^{\infty}(\theta)]^T$  is diagonal for all $\theta\in\Theta$,  with $H^\infty(\theta)\triangleq\lim_{z\rightarrow\infty}H(z,\theta)$.
\end{enumerate}
\end{assumption}

This assumption ensures that the spectral factorization of $\Phi_{\bar{v}}$ admits a unique spectral factor $(I-G)^{-1}H$ \citep{HarmAutomatica}, i.e., the mapping from the white noises to the internal signals as in \eqref{eq:EtoImap}. Then we can reformulate Definition~\ref{def:defnOriIden} by considering the uniqueness of network modules given $T_{\mathcal{W}\mathcal{X}} = [T_{\mathcal{W}\mathcal{R}} \ \ (I-G)^{-1} H]$.

\begin{proposition} \label{eq:finalDefn}
Given a network model set $\mathcal{M}$ that satisfies Assumption~\ref{ass:feedthrough}, consider any subset $\bar{\mathcal{W}}_j \subseteq \mathcal{W}_j$, $\theta_0 \in \Theta$ and the following implication:
\begin{align*}
T_{\mathcal{W} \mathcal{X}}(q,\theta_0)\! =\! T_{\mathcal{W} \mathcal{X}}(q,\theta_1)
\Rightarrow
\!G_{j\bar{\mathcal{W}}_j}(q,\theta_0)\! =\!  G_{j\bar{\mathcal{W}}_j}(q,\theta_1),
\end{align*}
for all $\theta_1 \in \Theta$. Then $G_{j\bar{\mathcal{W}}_j}(q,\theta)$ is globally (generically) identifiable in $\mathcal{M}$ from $(w,r)$ if and only if the above implication holds for all (almost all) $\theta_0 \in \Theta$.
%
\end{proposition}
\begin{pf}
This result is a direct consequence of Propositions 1 and 2 in \citep{HarmAutomatica}. $\hfill{} \blacksquare$
\end{pf}

In contrast to Definition~\ref{def:defnOriIden}, Proposition~\ref{eq:finalDefn} concerns the mapping $T_{\mathcal{W}\mathcal{X}}$ from both $r$ and $e$ signals to the internal signals. Therefore, the $r$ and $e$ signals play the same role in the identifiability analysis. This is also a key difference from the identifiability concept in \cite{bazanella2017identifiability, hendrickx2018identifiability}, where the noise information is not exploited. Proposition~\ref{eq:finalDefn} also extends trivially to the single module case where $G_{j\bar{\mathcal{W}}_j}$ only contains one entry, and to the full network case where $\bar{\mathcal{W}}_j = \mathcal{W}_j$ and $j$ varies over all $j \in [1,L]$.

\subsection{Problem formulation}
The identifiability issue is first illustrated in Example~\ref{exam:Problem}.

\begin{example} \label{exam:Problem}
\begin{figure}[h]
\begin{minipage}{0.23\textwidth}
\centering
\includegraphics[scale=0.3]{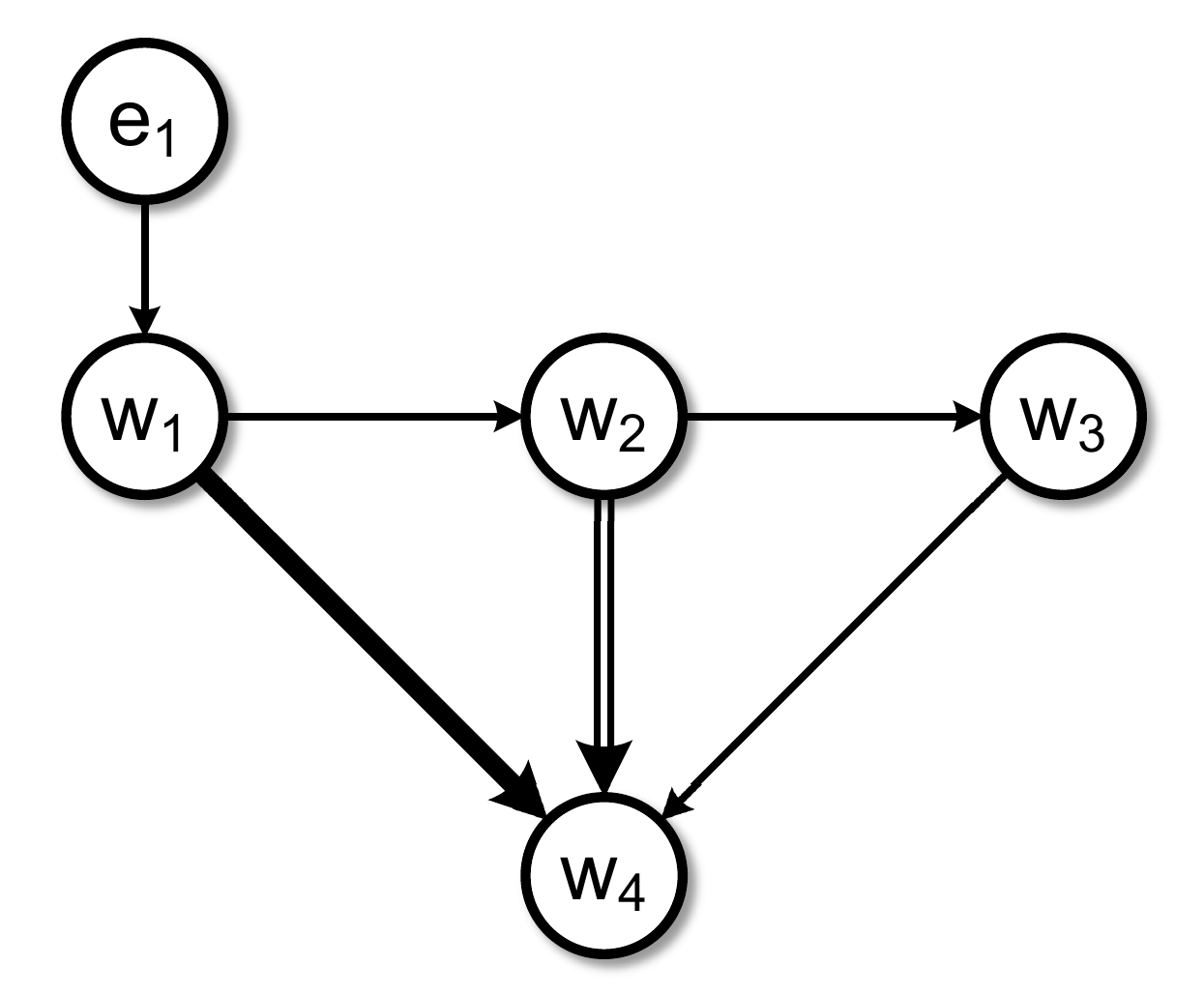}
\\(a)
\end{minipage}
\begin{minipage}{0.23\textwidth}
\centering
\includegraphics[scale=0.3]{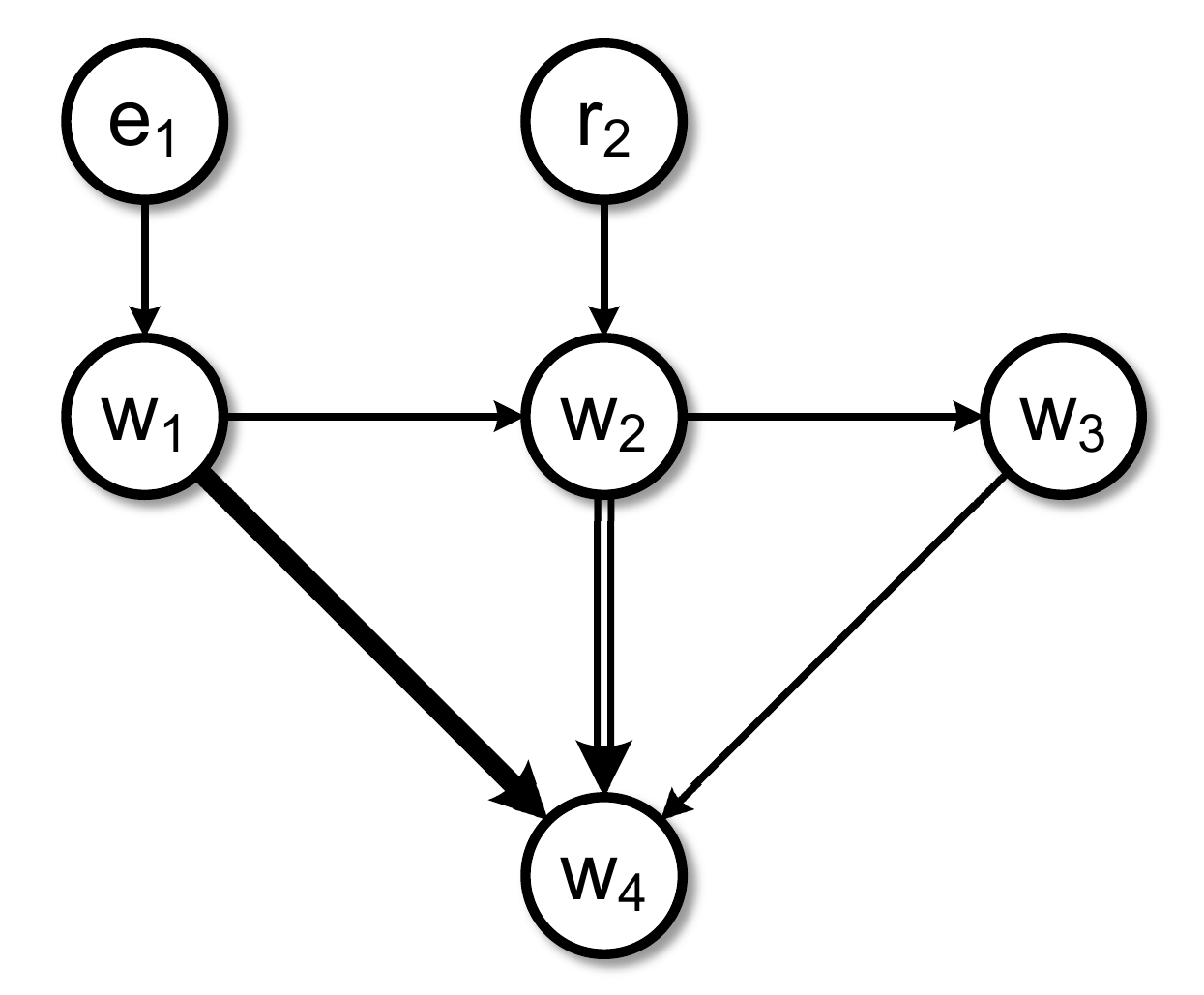}
\\(b)
\end{minipage}
\caption{Generic identifiability of $G_{41}$ is considered (thick line), and $G^0_{42}$ is known (double-line edge). $G_{41}$ is not generically identifiable in (a) but becomes generically identifiable in (b) if an extra signal $r_2$ is allocated at~$w_2$.}
\label{fig:implication}
\end{figure}
Consider a network model set whose graph is shown in Fig.~\ref{fig:implication}(a), where there is one white noise $e_1$ and four measured internal signals. All edges are parametrized except for $G_{42}$ which is fixed to $G_{42}^0$. The identifiability of module $G_{41}$ is considered, and recall that the dependence of transfer functions on $q$ and $\theta$ is omitted for simplicity of notation. Then based on Proposition~\ref{eq:finalDefn}, the identifiability of $G_{41}$ concerns the uniqueness of $G_{41}$ given the mapping $T_{\{w_1,w_2,w_3,w_4\}\{e_1\}}$ from $e_1$ to all the internal signals. According to Fig.~\ref{fig:implication}(a), the mapping satisfies that $T_{\{w_1\}\{e_1\}} = H_{11}$, $T_{\{w_2\}\{e_1\}} = G_{21}H_{11}$, $T_{\{w_3\}\{e_1\}} = G_{32}G_{21}H_{11}$ and $$T_{\{w_4\}\{e_1\}} = H_{11}(G_{41}+ G^0_{42} G_{21}+G_{43}G_{32}G_{21}),
$$
where $G^0_{42}$ and the submatrices of $T_{\{w_1,w_2,w_3,w_4\}\{e_1\}}$ are given, but $H_{11}$ and the other entries in $G$ are unknown. The above equations further lead to
\begin{equation*}
T_{\{w_4\}\{e_1\}} = T_{\{w_1\}\{e_1\}}G_{41}+G^0_{42}T_{\{w_2\}\{e_1\}}+G_{43}T_{\{w_3\}\{e_1\}},
\end{equation*}
which is not sufficient for a unique solution of $G_{41}$ due to the two unknowns $G_{41}$ and $G_{43}$. Therefore, $G_{41}$ is neither globally nor generically identifiable in Fig.~\ref{fig:implication}(a).

However, if an additional signal $r_2$ is allocated at $w_2$ as in Fig.~\ref{fig:implication}(b), additional mappings from $r_2$ to the measured internal signals are given. Then it can be found that
$$
G_{41} = ( T_{\{w_4\}\{e_1\}}- \frac{T_{\{w_2\}\{e_1\}} T_{\{w_4\}\{r_2\}} }{T_{\{w_2\}\{r_2\}}}  )/ T_{\{w_1\}\{e_1\}},
$$
which shows that a unique $G_{41}$ can be found if the above inverses exist. Recall that the mappings depend on parameter $\theta$, and thus $G_{41}$ is globally (generically) identifiable if the inverses exist for all (almost all) $\theta \in \Theta$.
\end{example}
As shown in the above example, identifiability is related to the uniqueness of modules given the mapping $T_{\mathcal{W}\mathcal{X}}$. In addition, it is possible to achieve identifiability by allocating extra excitation signals in the network.

Given a model set and its associated graph, we first develop sufficient and necessary conditions for the generic identifiability of a subnetwork $G_{j\bar{\mathcal{W}}_j}$ when all the internal signals are measured. More importantly, to avoid the complex algebraic verification as in Example~\ref{exam:Problem}, we aim to develop graphical conditions that can be verified efficiently by only inspecting the network topology. Furthermore, when a subnetwork is not generically identifiable,  we aim to develop graphical synthesis approaches that allocate additional excitation signals to  automatically achieve the generic identifiability of a subnetwork.

\subsection{Notations and definitions}
\label{sec:notation}
The following notations are used throughout the paper.
Matrix $T_{\bar{\mathcal{W}}  \bar{\mathcal{X}}}(q,\theta)$ is called having generically full rank if it has full rank for almost all $\theta$. More generally, a property that depends on $\theta$ is said to hold generically if it holds for almost all $\theta$. A matrix $C$ is called a selection matrix if it consists of a subset of rows of an identity matrix.   \\
In a graph $\mathcal{G}$, for a directed edge $(w_i,w_j)$, $w_i$ is called an \textit{in-neighbor} of $w_j$, and $w_j$ is an \textit{out-neighbor} of $w_i$. The set $\mathcal{N}^+_{\bar{\mathcal{V}}}$ contains all out-neightbors of the vertices in $\bar{\mathcal{V}}$, and the set $\mathcal{N}^-_{\bar{\mathcal{V}}}$ all in-neighbors. A (directed) \textit{path} from $w_i$ to $w_j$ is a sequence of vertices and out-going edges starting from $w_i$ to $w_j$ without repeating any vertex, and $w_j$ is said \textit{reachable} by $w_i$.  A single vertex is also regarded as a directed path to itself. \textit{Internal vertices} of a path are the vertices excluding the starting and the ending vertices.\\
Two directed paths are called \textit{vertex disjoint} if they do not share any vertex, including the starting and ending vertices, otherwise they \textit{intersect}. Given two subsets of vertices $\mathcal{V}_1$ and $\mathcal{V}_2$, $b_{\mathcal{V}_1 \to \mathcal{V}_2}$ denotes the maximum number of vertex disjoint paths from $\mathcal{V}_1$ to $\mathcal{V}_2$. A vertex set $\mathcal{D}$ is a $\mathcal{V}_1 - \mathcal{V}_2$ \textit{disconnecting set} if it intersects with all paths from $\mathcal{V}_1$ to $\mathcal{V}_2$, and it is a minimum disconnecting set if it has the minimum cardinally among all $\mathcal{V}_1-\mathcal{V}_2$ disconnecting sets \citep{schrijver2003combinatorial}. Note that $\mathcal{D}$ may also include vertices in $\mathcal{V}_1 \cup \mathcal{V}_2$.
\begin{lemma}\label{lem0}\citep{hendrickx2018identifiability}
For a directed graph and given a $\mathcal{V}_1 - \mathcal{V}_2$ disconnecting set $\mathcal{D}$, consider the division of all the vertices $\mathcal{V}$ into three disjoint sets $\mathcal{S} \cup \mathcal{D} \cup \mathcal{P}$ as follows: set $\mathcal{S}$ contains all vertices reachable by $\mathcal{V}_1$ without intersecting $\mathcal{D}$, and $\mathcal{P} = \mathcal{V} \setminus (\mathcal{D} \cup \mathcal{S})$. Then it holds that no directed edge exists from $\mathcal{S}$ to $\mathcal{P}$.
\end{lemma}

The duality between vertex disjoint paths and disconnecting sets is explained in \textit{Menger's theorem}.
\begin{theorem} \citep{schrijver2003combinatorial} \label{thm:Menger}
Let $\mathcal{V}_1$, $\mathcal{V}_2$ be two subsets of the vertices in a directed graph. The maximum number of vertex disjoint paths from $\mathcal{V}_1$ to $\mathcal{V}_2$ equals the cardinality of a minimum disconnecting set from $\mathcal{V}_1$ to $\mathcal{V}_2$.
\end{theorem}

\section{Algebraic conditions for identifiability} \label{sec:Algebra}
We first investigate under which conditions a subnetwork $G_{j\bar{\mathcal{W}}_j}$ is globally and generically identifiable. An algebraic rank condition is first presented in this section which will serve as a basis for deriving a more attractive graphical condition in the next section. Recall \eqref{eq:EtoImap} which leads to
$
(I-G)T_{\mathcal{W} \mathcal{X}} = X.
$
It then follows from Proposition~\ref{eq:finalDefn} that identifiability essentially reflects the unique solutions of modules in $G$ from a given $T_{\mathcal{W}\mathcal{X}}$ and thus is related to the rank of $T_{\mathcal{W} \mathcal{X}}$. We first define the following set related to the output $w_j$:
\begin{itemize}
    \item  $\mathcal{X}_j$: the set of all the external signals that do not have any unknown edge to $w_j$.
\end{itemize}
Then the following sufficient conditions for identifiability can be obtained from Theorem~V.1 in \citep{hendrickx2018identifiability} analogously.

\begin{lemma} \label{theorem:rank}
Given a model set $\mathcal{M} $ that satisfies Assumption~\ref{ass:feedthrough}. Then $G_{j \bar{\mathcal{W}}_j}(q,\theta)$ is globally (generically) identifiable in $\mathcal{M}$ from $(w,r)$ if the following conditions hold:
\begin{subequations} \label{eq:rankCondi}
\begin{align}
 \rank[T_{\bar{\mathcal{W}}_j \mathcal{X}_j}(q,\theta)] = &\ |\bar{\mathcal{W}}_j|,\ \mbox{and}\label{eq:6a}\\
\rank[T_{\mathcal{W}_j \mathcal{X}_j}(q,\theta)] = &\ \rank[T_{\bar{\mathcal{W}}_j \mathcal{X}_j}(q,\theta)] \nonumber \\ &\ + \rank[T_{(\mathcal{W}_j \setminus \bar{\mathcal{W}}_j) \mathcal{X}_j}(q,\theta)]\label{eq:6b}
\end{align}
\end{subequations}
for all (almost all) $\theta \in \Theta$.
\end{lemma}

The above result shows that identifiability of $G_{j\bar{\mathcal{W}}_j}$ holds if (i) the mapping from $\mathcal{X}_j$ to the inputs of the target modules has full row rank, and (ii) the rows in the above mapping and the ones in the mapping from $\mathcal{X}_j$ to the other inputs of $w_j$ are linearly independent. Here, we consider the signals in $\mathcal{X}_j$ instead of $\mathcal{X}$, because the ones in $\mathcal{X} \setminus \mathcal{X}_j$, which have unknown edges to $w_j$, contribute to extra unknowns in the analysis in the form of a parametrized disturbance model for $v_j$, and thus are not helpful for the identifiability analysis. If the corresponding noise model is fixed/known, the corresponding white noise source becomes an element of $\mathcal{X}_j$ and does act as an excitation source in the identifiability analysis.

The sufficient conditions in Lemma \ref{theorem:rank} are implied in the proof of Theorem~V.1 of \citep{hendrickx2018identifiability}. While in that proof they were shown to be also necessary, that necessity proof cannot be applied directly to our current setting, owing to a different definition of generic identifiability:  Definition~\ref{def:defnOriIden} is based on a particular model set, while the identifiability concept in \cite{hendrickx2018identifiability} concerns a single network model and its associated graph. To establish the necessity of the rank conditions in this work, we first introduce two technical conditions on the model set $\mathcal{M}$.
\begin{assumption} \label{ass:indePara}
All the parametrized entries in $M(\theta)$ are parametrized independently.
\end{assumption}

\begin{assumption} \label{ass:openset}
Given the network model set $\mathcal{M}$ and the corresponding $\mathbb{G}^\star$ defined in Definition~\ref{def:strucModelSet}, $\{G(q,\theta)| \theta \in \Theta\}$ is a non-empty open subset\footnote{Openness is considered in the metric space equipped with the $H_\infty$ norm.}of $\mathbb{G}^\star$.
\end{assumption}

Then the necessity of the rank conditions for identifiability can be obtained as follows.
\begin{theorem}
\label{theorem:Goodrank}
Given a model set $\mathcal{M} $ that satisfies Assumption~\ref{ass:feedthrough}, if $\mathcal{M}$ also satisfies Assumptions~\ref{ass:indePara} and \ref{ass:openset}, conditions (\ref{eq:6a})-(\ref{eq:6b}) in Lemma \ref{theorem:rank} are also necessary conditions for the results of the lemma.
\end{theorem}
The proof of the above result is presented in Appendix. Theorem~\ref{theorem:Goodrank} extends the identifiability conditions in \citep{HarmAutomatica,HarmGraphCDC} from a full network and a single module to any subnetwork, such that the full network and the single module identifiability become special cases. In addition, with extra technical efforts in the analysis of Theorem~\ref{theorem:Goodrank}, Assumption~\ref{ass:openset} becomes less conservative than the assumption in \citep{HarmAutomatica,HarmGraphCDC} for the necessity of identifiability.

The rank tests in Theorem~\ref{theorem:Goodrank} may require to invert $(I-G)$ for computing the external-to-internal mappings, which can be computationally expensive when the network is large-scale. The following result shows that the rank of any submatrix of $T_{\mathcal{W}\mathcal{X}}$ can be also evaluated without inverting the matrix $I-G$ for computing $T_{\mathcal{W}\mathcal{X}}$.

\begin{lemma} \label{lem:Fintro}
Consider a dynamic network and any subsets $\bar{\mathcal{W}} \subseteq \mathcal{W}$ and $\bar{\mathcal{X}} \subseteq \mathcal{X}$. Define the matrix $F(\bar{\mathcal{W}},\bar{\mathcal{X}})$ as
\begin{equation}
F(\bar{\mathcal{W}},\bar{\mathcal{X}}) \triangleq \begin{bmatrix}
[G(q)-I]_{\mathcal{W} (\mathcal{W} \setminus \bar{\mathcal{W}})} &  X_{\mathcal{W} \bar{\mathcal{X}}}(q)
\end{bmatrix}. \label{eq:Fdfn}
\end{equation}
Then it holds that
$
\rank(T_{\bar{\mathcal{W}}\bar{\mathcal{X}}}(q)) = \rank[F(\bar{\mathcal{W}},\bar{\mathcal{X}})] + |\bar{\mathcal{W}}|- L,
$
where $L$ is the total number of internal signals.
\end{lemma}
\begin{pf}
Recall that $T_{\bar{\mathcal{W}}\bar{\mathcal{X}}} = C (I-G)^{-1}\bar{X}$, where $\bar{X}= X_{\mathcal{W} \bar{\mathcal{X}}}$ and  $C$ is a selection matrix that extracts the rows corresponding to $\bar{\mathcal{W}}$. Define
\begin{equation}
\bar{F} = \begin{bmatrix}
G-I &  \bar{X} \\
C & 0
\end{bmatrix}, \label{eq:F}
\end{equation} then according to \citep{van1991graph}, $\bar{F}$ can be written as
$$
\begin{bmatrix}
I & 0\\
C(G\! -\! I)^{-1}& I
\end{bmatrix}\!\! \begin{bmatrix}
G-I& 0 \\
0 & C(I\! -\! G)^{-1}\bar{X}
\end{bmatrix}\!\! \begin{bmatrix}
I & (G\! -\! I)^{-1}\bar{X}\\
0 & I
\end{bmatrix}.
$$
Thus, $\rank(\bar{F}) = \rank(T_{\bar{\mathcal{W}}\bar{\mathcal{X}}}) + L$. Since $F$ is obtained from $\bar F$ by removing the columns in $\bar F$ that correspond to the nonzero entries of $C$, which through the structure of $\bar F$ are linearly independent in $F$, it follows that $\rank (F) = \rank (\bar F) - |\bar{\mathcal{W}}|$, which proves the result. $\hfill{} \blacksquare$
\end{pf}

With the above result, the rank condition in Theorem~\ref{theorem:Goodrank} can be reformulated
as follows.

\begin{proposition}\label{prop:F}
Consider a network model set and the matrix $F$ defined in \eqref{eq:Fdfn}. The conditions (\ref{eq:6a})-(\ref{eq:6b}) in Theorem~\ref{theorem:Goodrank} are equivalently reformulated as $F(\bar{\mathcal{W}}_j,\mathcal{X}_j)$ having full row rank and
\begin{align*}
\rank[F(\mathcal{W}_j,\mathcal{X}_j)] =  &\rank[F(\bar{\mathcal{W}}_j,\mathcal{X}_j)] \\ &+ \rank[F( \mathcal{W}_j \setminus \bar{\mathcal{W}}_j,\mathcal{X}_j)]
-L,
\end{align*}
for all (almost all) $\theta \in \Theta$.
\end{proposition}

\section{Path-based conditions for generic identifiability}
\label{sec:AllocateSingleNew}
When analysing generic identifiability, the results of Lemma \ref{theorem:rank} and Proposition \ref{prop:F} show that the rank of the parametrized matrices have to be evaluated over almost all $\theta\in\Theta$. The resulting generic rank of transfer matrices in dynamic networks can be evaluated by determining the number of vertex disjoint paths between particular vertices in the graph induced by the model set, i.e.
\begin{equation} \label{eq:vdp}
 b_{\bar{\mathcal{X}} \to \bar{\mathcal{W}}} = \rank\ T_{\bar{\mathcal{W}} \bar{\mathcal{X}}}(q,\theta) \ \ \mbox{for almost all }\theta\in\Theta.
\end{equation}
This result from \citep{hendrickx2018identifiability} is very attractive as it allows us to avoid numerical rank evaluations, and therefore significantly facilitates the analysis of generic identifiability. However, the result in \citep{hendrickx2018identifiability} has been derived for the particular situation that $X=I$, and that all edges in the graph $\mathcal{G}$ are parametrized, i.e. without having fixed / known  modules in the model set. In this section we will extend the results beyond these limitations.
\\
The issue caused by known transfer functions is explained in the following example.
\begin{example} \label{exam:fixedmodule}
Consider two network model sets in Fig.~\ref{fig:genericIssue}, where the model set in (b) contains known modules. In addition, assume  $R_{11} = R_{22} = 1$ for simplicity.
Then in Fig.~\ref{fig:genericIssue}(a), $b_{\{r_1,r_2\} \to \{w_3,w_4\}} =2$, i.e. there exist maximally two paths $r_1 \to w_1 \to w_3$ and $r_2 \to w_2 \to w_4$ that are vertex disjoint. This implies that $\rank(T_{\{w_3,w_4\}\{r_1,r_2\}})= 2$ generically according to \eqref{eq:vdp}, which can also be seen from
\begin{equation}
\det(T_{\{w_3,w_4\}\{r_1,r_2\}}) = G_{31}G_{42} -G_{41}G_{32}. \label{eq:fixedModuleExample}
\end{equation}
As each module is an analytic function of independent parameters, the determinant in \eqref{eq:fixedModuleExample} is a non-constant analytic function of $\theta$ and thus is non-zero for almost all $\theta$. This matches with \eqref{eq:vdp} that $rank(T_{\{w_3,w_4\}\{r_1,r_2\}}) = b_{\{r_1,r_2\} \to \{w_3,w_4\}} =2$ generically. \\
However, $T_{\{w_3,w_4\}\{r_1,r_2\}}$ is not of full rank in Fig.~\ref{fig:genericIssue}(b) if the known modules $G^0_{31}, G^0_{42}, G^0_{41}, G^0_{32}$ take values such that
$G^0_{31}G^0_{42} -G^0_{41}G^0_{32}=0$. Then $\det(T_{\{w_3,w_4\}\{r_1,r_2\}}) =0$ for any $\theta$ and thus \eqref{eq:vdp} does not hold.
%
\begin{figure}[t]
\begin{minipage}{0.23\textwidth}
\centering
\includegraphics[scale=0.3]{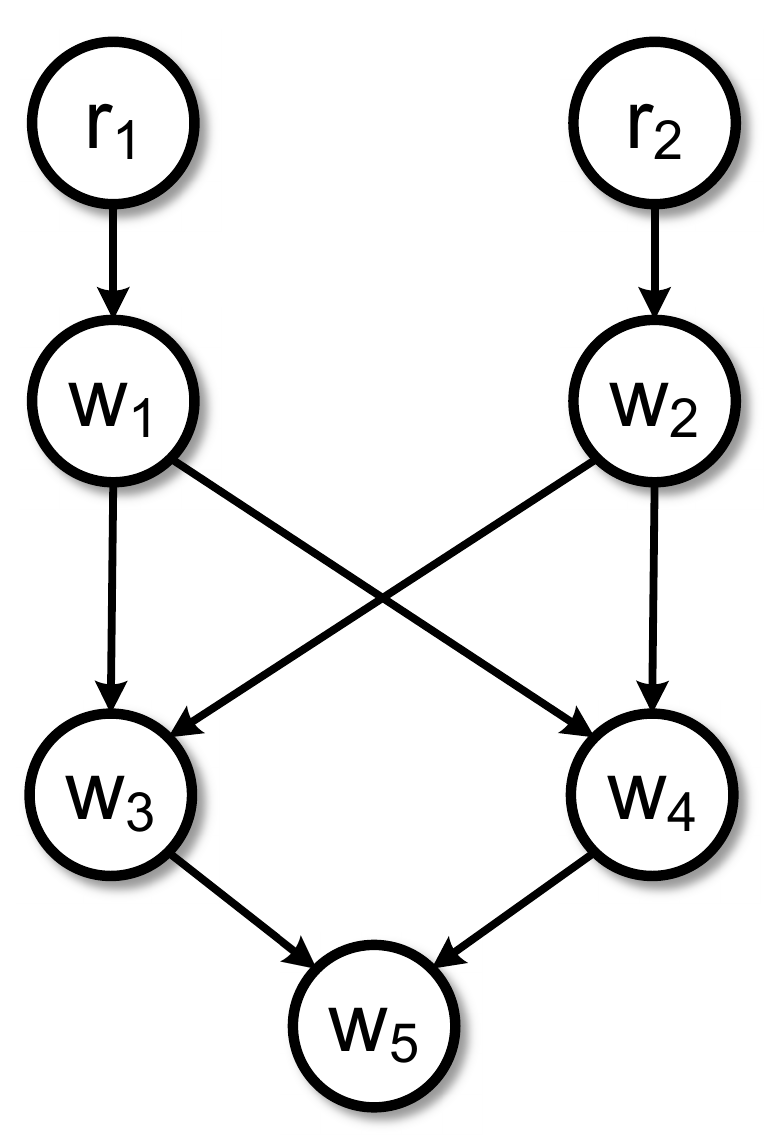}
\\(a)
\end{minipage}
\begin{minipage}{0.23\textwidth}
\centering
\includegraphics[scale=0.3]{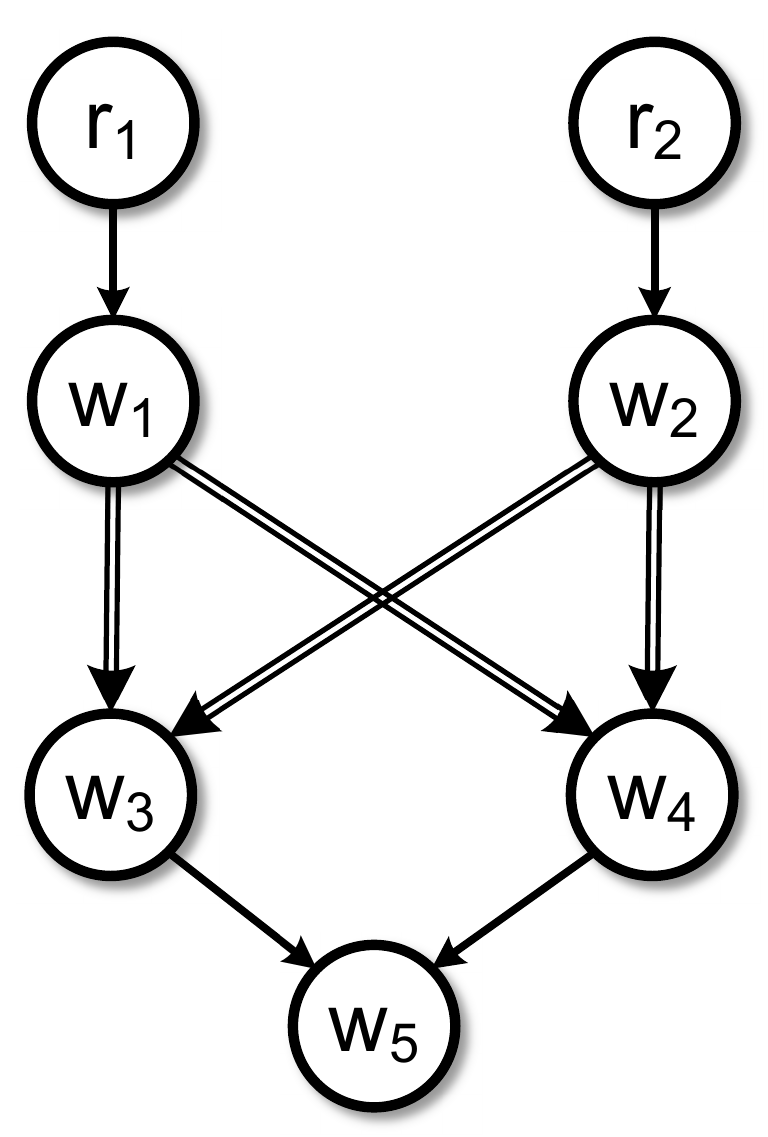}
\\(b)
\end{minipage}
\caption{Two network model sets where (b) contains known modules (double-line edges).}
\label{fig:genericIssue}
\end{figure}
\hfill{} $\blacksquare$
\end{example}
As shown in Example~\ref{exam:fixedmodule}, the presence of known modules may cause the equality in \eqref{eq:vdp} to fail. In order to circumvent this, we introduce an additional condition  based on the concept of structural rank.
\begin{defn}[\citep{steffen2005control}]
The structural rank of a matrix is the maximum rank of all matrices with the same nonzero pattern. A matrix has full structural rank if it can be permuted so that the diagonal has no zero entries.
\end{defn}
The property of structural full rank depends solely on the sparsity pattern of the matrix and does not rely on the numerical values of the entries.
%
In order to characterize and exclude the situations as presented in Example \ref{exam:fixedmodule}, we introduce the following assumption.
\begin{assumption}\label{ass5}
In model set $\mathcal{M}$, the rank of any submatrix of $[(G(q,\theta)-I) \text{ } X(q,\theta)]$, which does not depend on $\theta$, is equal to its structural rank. $\hfill{} \blacksquare$
\end{assumption}

The above assumption prevents known entries to be chosen in such a specific manner that they cause a rank deficiency of $T_{\bar{\mathcal{W}}\bar{\mathcal{X}}}$, which is not captured by the number of vertex disjoint paths. Using this assumption, equality \eqref{eq:vdp} can be extended
from \cite{hendrickx2018identifiability} to incorporate known transfer operators in the model set, as shown in the following result.

\begin{proposition} \label{prop:TsubRank}
Consider a model set $\mathcal{M} $ with graph $\mathcal{G}$ that satisfies Assumptions~\ref{ass:indePara} and \ref{ass5}. Then for any subsets $\bar{\mathcal{W}} \subseteq \mathcal{W}$ and $\bar{\mathcal{X}} \subseteq \mathcal{X}$, it holds that
$$
b_{\bar{\mathcal{X}} \to \bar{\mathcal{W}}} = \rank[T_{\bar{\mathcal{W}}\bar{\mathcal{X}}}(q,\theta)] \ \ \mbox{for almost all } \theta \in \Theta.
$$
\end{proposition}
The proof of the above result is presented in Appendix. By combining Proposition~\ref{prop:TsubRank} with Theorem~\ref{theorem:Goodrank}, the graphical condition for generic identifiability can be obtained, that follows immediately from the previous results:
\begin{theorem} \label{theorem:PathCondi}
Consider a model set $\mathcal{M}$ that satisfies Assumptions~\ref{ass:feedthrough}, \ref{ass:indePara} and \ref{ass5}. Then $G_{j\bar{\mathcal{W}}_j}(q,\theta)$ is generically identifiable in $\mathcal{M}$ from $(w,r)$ if the following conditions hold:
\begin{subequations}
\begin{align}
 & b_{\mathcal{X}_j \to \bar{\mathcal{W}}_j } = | \bar{\mathcal{W}}_j |, \ \mbox{and} \label{eq1112a}\\
 & b_{\mathcal{X}_j \to \mathcal{W}_j } = b_{\mathcal{X}_j \to \bar{\mathcal{W}}_j} + b_{\mathcal{X}_j \to \mathcal{W}_j \setminus \bar{\mathcal{W}}_j}. \label{eq1112b}
\end{align}
\end{subequations}
When $\mathcal{M}$ additionally satisfies Assumption~\ref{ass:openset}, the above conditions are also necessary.
\end{theorem}
The above result shows that generic identifiability of the target modules in $G_{j\bar{\mathcal{W}}_j}$ is ensured if (i) the inputs $\bar{\mathcal{W}}_j$ of the target modules are excited by $\mathcal{X}_j$ in the sense that $\mathcal{X}_j$ has a sufficient number of vertex disjoint paths to these inputs, and (ii) the inputs $\bar{\mathcal{W}}_j$ receive excitation that is independent of the excitation for the remaining inputs $\mathcal{W}_j \setminus \bar{\mathcal{W}}_j$. The result of Theorem~\ref{theorem:PathCondi} is also illustrated in Example ~\ref{examplePath}.
An important special case of Theorem~\ref{theorem:PathCondi} is when $\bar{\mathcal{W}}_j = \mathcal{W}_j$, i.e. all parametrized modules that map into $w_j$ are in the subnetwork under study. In this case, conditions \eqref{eq1112a} and \eqref{eq1112b} can be simplified into a single condition as
$
b_{\mathcal{X}_j \to \mathcal{W}_j} = |\mathcal{W}_j|.
$
\begin{example} \label{examplePath}
The generic identifiability of $G_{41}$ in the network model set in Fig.~\ref{fig:implication}(a) can be verified using Theorem~\ref{theorem:PathCondi} as follows. In this model set, $\mathcal{X}_4 = \{e_1\}$, $\mathcal{W}_4 = \{w_1, w_3\}$ and $\bar{\mathcal{W}}_4 = \{w_1\}$. As there exists a path from $e_1$ to $w_1$, we have $b_{\mathcal{X}_4 \to \bar{\mathcal{W}}_4 } = 1 = |\bar{\mathcal{W}}_4|$ and thus \eqref{eq1112a} is satisfied. In addition, since all paths from $e_1$ to $\{w_1, w_3\}$ intersect with $w_1$, we have $b_{\mathcal{X}_4 \to \mathcal{W}_4 } =b_{\mathcal{X}_4 \to \mathcal{W}_4\setminus  \bar{\mathcal{W}}_4} = 1$, which shows that
$
b_{\mathcal{X}_4 \to \mathcal{W}_4 } < b_{\mathcal{X}_4 \to \bar{\mathcal{W}}_4} + b_{\mathcal{X}_4 \to \mathcal{W}_4 \setminus \bar{\mathcal{W}}_4},
$
and thus \eqref{eq1112b} is not satisfied, which shows that $G_{41}$ is not generically identifiable in Fig.~\ref{fig:implication}(a).
\end{example}

Compared to Theorem~\ref{theorem:Goodrank} and the analysis in Example~\ref{exam:Problem}, the conditions in Theorem~\ref{theorem:PathCondi} are fully graphical and thus can be evaluated more efficiently by only inspecting the network topology. Standard graphical algorithms are available for computing the maximum number of vertex disjoint paths \citep{schrijver2003combinatorial}. Note that the conditions (\ref{eq1112a})-(\ref{eq1112b}) are similar to the ones in \cite{hendrickx2018identifiability}, but now extended to a different situation. Besides the incorporation of known modules and the model-set-based identifiability in Definition~\ref{def:defnOriIden}, the set $\mathcal{X}_j$ contains both measured excitation signals and unmeasured white noises, instead of only $r$ signals as considered in \cite{hendrickx2018identifiability}. Thus, to satisfy the graphical conditions in the theorem, the white noises can compensate for a possible lack of excitation signals.

%
%
\section{Disconnecting-set-based conditions}\label{sec:disconset}
\subsection{Generic identifiability based on disconnecting sets}
The graphic condition in Theorem~\ref{theorem:PathCondi} can be applied to analyzing generic identifiability in a given model set. However, it does not explicitly indicate where to allocate external signals such that a particular set of modules becomes generically identifiable. To solve this synthesis question, a new analytic result is developed in this section by exploring the duality between vertex disjoint paths and disconnecting sets, as also indicated in \cite{hendrickx2018identifiability}. In this section, we will exploit this relationship in a novel way, by suitably exploiting it for the situation of subnetwork generic identifiability, while providing a solution to the synthesis question as well. This synthesis question will be addressed in more detail in Section~\ref{sec:synthesis}.

A set of vertices is called a disconnecting set from vertex set $\mathcal{V}_1$ to $\mathcal{V}_2$ if all paths from $\mathcal{V}_1$ to $\mathcal{V}_2$ pass through the set, see Section~\ref{sec:notation}. Its
relevance in the analysis of generic identifiability is illustrated in Example~\ref{example1}.
\begin{example}
Starting from Fig.~\ref{fig:implication}(a), if $r_2$ is allocated at $w_2$ as in Fig.~\ref{fig:implication}(b) such that $\mathcal{X}_4 = \{e_1,r_2\}$ and $b_{\mathcal{X}_4 \to \mathcal{W}_4 }=2$, we have $b_{\mathcal{X}_4 \to \mathcal{W}_4 } = b_{\mathcal{X}_4 \to \bar{\mathcal{W}}_4 } + b_{\mathcal{X}_4 \to \mathcal{W}_4\setminus  \bar{\mathcal{W}}_4}=2$, which shows that $G_{41}$ becomes generically identifiable in Fig.~\ref{fig:implication}(b) based on Theorem~\ref{theorem:PathCondi}. It is important to note that the vertex $w_2$, where the new $r_2$ is allocated, is a disconnecting set from the target input $w_1$ to the other inputs of $w_4$.
\label{example1}
$\hfill{} \blacksquare$
\end{example}

It is observed from the above example that when $G_{ji}$ is not generically identifiable, its identifiability can be achieved when the vertices in a disconnecting set from $\{w_i\}$ to the other in-neighbors of $w_j$ are excited. This observation is generalized to obtain a new identifiability condition based on disconnecting sets.\\

\begin{theorem} \label{theorem:SingleCutNew}
Consider the situation of Theorem \ref{theorem:PathCondi}. The conditions (\ref{eq1112a})-(\ref{eq1112b}) can equivalently be written as either one of the following two equivalent conditions:
\begin{enumerate}[label=(\alph*)]
\item \label{item1}There exists a $\mathcal{X}_j - \mathcal{W}_j \setminus \bar{\mathcal{W}}_j$ disconnecting set $\mathcal{D}$ such that $b_{\mathcal{X}_j  \to \bar{\mathcal{W}}_j \cup \mathcal{D}} = |\mathcal{D}| + |\bar{\mathcal{W}}_j|.$
\item \label{item2}There exists a set of external signals $\bar{\mathcal{X}}_j \subseteq \mathcal{X}_j$ and a $\bar{\mathcal{X}}_j - \mathcal{W}_j \setminus \bar{\mathcal{W}}_j$ disconnecting set $\mathcal{D}$ such that
\begin{equation}
b_{\bar{\mathcal{X}}_j  \to \bar{\mathcal{W}}_j \cup \mathcal{D}} = |\mathcal{D}| + |\bar{\mathcal{W}}_j|. \label{eq:TheoremCut}
\end{equation}
\end{enumerate}
\end{theorem}
The proof of the above result is presented in Appendix. As visualized in Fig.~\ref{fig:thmVisu}, this result states that the target modules in $G_{j \bar{\mathcal{W}}_j}$ are generically identifiable, if the inputs of the target modules in $\bar{\mathcal{W}}_j$ and the signals in $\mathcal{D}$ are excited by external signals through vertex disjoint paths, where $\mathcal{D}$ contains the vertices that block the paths from the target inputs to the other inputs that have unknown edges to $w_j$. In contrast to Theorem~\ref{theorem:PathCondi}, the conditions in Theorem~\ref{theorem:SingleCutNew} explicitly state that the signals in $\bar{\mathcal{W}}_j \cup \mathcal{D}$ should be excited. When the target modules are not identifiable, the above result will be exploited in Section~\ref{sec:synthesis} to allocate additional excitation signals for achieving identifiability.

In Theorem~\ref{theorem:SingleCutNew}, condition~\ref{item1} more closely resembles Theorem~\ref{theorem:PathCondi} as it involves all external signals in $\mathcal{X}_j$; on the other hand, the condition~\ref{item2} considers any subset of signals in $\mathcal{X}_j$, and this ability of considering subsets of $\mathcal{X}_j$ will be exploited for the design of synthesis approaches.
\begin{figure}[h]
\centering
\includegraphics[scale=0.24]{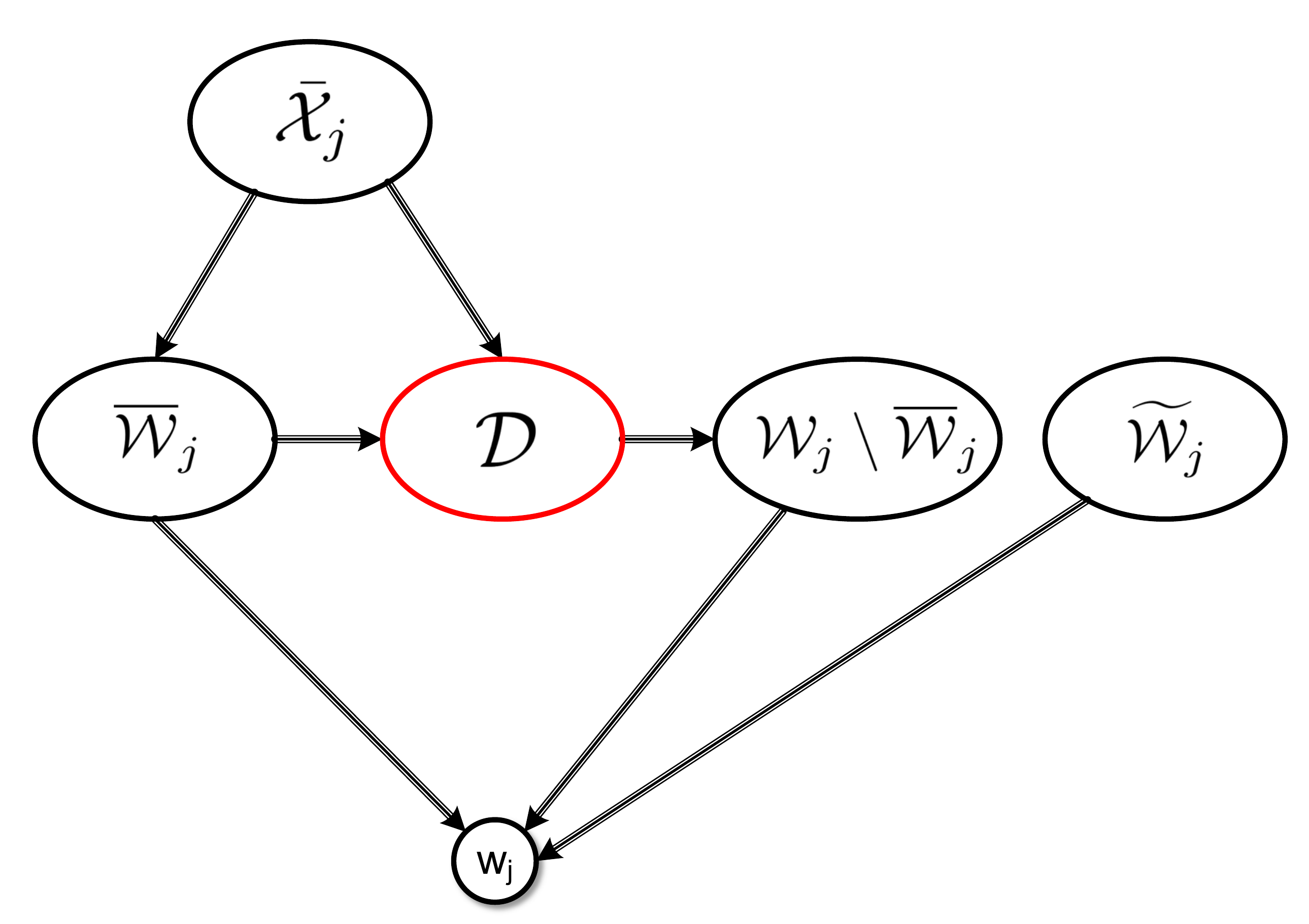}
\caption{Visualization of condition \ref{item2} in Theorem~\ref{theorem:SingleCutNew}, where $\tilde{\mathcal{W}}_j$ collects the internal signals that have known edges to $w_j$.} \label{fig:thmVisu}
\end{figure}

\subsection{Algebraic interpretation of disconnecting sets} \label{sec:algebraDis}
A disconnecting set that disconnects external signals from internal signals in the graph of a network model set, gives rise to an attractive factorization of the external-to-internal mapping for the associated dynamic network.

\begin{theorem} \label{the:Decompo2}
Consider a network model $M$ with a disconnecting set $\mathcal{D} \subseteq \mathcal{V}$ from $\bar{\mathcal{X}} \subseteq \mathcal{X}$ to $\bar{\mathcal{W}} \subseteq \mathcal{W}$. Then there exists a proper transfer matrix $K(q)$ such that
\begin{equation}
T_{\bar{\mathcal{W}}  \bar{\mathcal{X}} }(q) = K(q) T_{\mathcal{D} \bar{\mathcal{X}}}(q).  \label{eq:11a}
\end{equation}
\end{theorem}
The proof of the above result is presented in Appendix. The decomposition in \eqref{eq:11a} means that if all paths from $\bar{\mathcal{X}}$ to $\bar{\mathcal{W}}$ intersect with $\mathcal{D}$, then the signals in $\mathcal{D}$ act as auxiliary signals that contain all information from the signals in $\bar{\mathcal{X}}$ that is relevant for the signals in $\bar{\mathcal{W}}$. The explicit form of the $K$ matrix is specified in \eqref{eq:finalK}. The factorization in Theorem~\ref{the:Decompo2} will serve as a basis for formulating an indirect identification method for identifying $G_{j\bar{\mathcal{W}_j}}$. This will be addressed in Section \ref{sec:indirect}.

\subsection{Relation with the parallel path and loop condition} \label{sec:RaraPath}
In dynamic network identification, there is an important condition used to select measured internal signals for identifying a single module $G_{ji}$. The condition was first formulated in \citep{dankers2015} as the {\it parallel path and loop condition} and selects a set of internal signals as follows. Determine a set of internal signals $\mathcal{D}$ according to:
\begin{itemize}
\item Every parallel path from $w_i$ to $w_j$, i.e. a path that is not the edge $G_{ji}$, passes through an internal vertex in $\mathcal{D}$, and
\item Every loop through $w_j$ passes through a vertex in $\mathcal{D}$.
\end{itemize}
If in a network the signals $w_i$, $w_j$ and $\mathcal{D}$ are retained, and all other internal signals are eliminated (immersed), e.g., because they can not be measured, then in the immersed network that leaves the measured internal signals invariant, the target module $G_{ji}$ remains invariant too. As a result, this condition has become an important tool for selecting internal signals to be measured for identification of a single module, see e.g. also \cite{weerts2019abstractions,ramaswamy2019local}. It can be shown that the above condition has a very strong link to disconnecting sets and Theorem~\ref{theorem:SingleCutNew}.

\begin{proposition} \label{propo:parallel}
Consider a model set where all non-zero modules in $G(q,\theta)$ are parametrized. Consider the module $G_{ji}$ and a set of internal signals $\mathcal{D}$ with $\{w_i\} \notin \mathcal{D}$. Then $\mathcal{D}$ is an $\{w_i\} - \mathcal{W}_j \setminus \{w_i\}$ disconnecting set if and only if $\mathcal{D}$ contains an internal vertex of every parallel path from $w_i$ to $w_j$ and a vertex of every loop around $w_j$.
\end{proposition}
The proof of the above result is presented in Appendix. While the parallel path and loop condition is used to select internal signals for identifying $G_{ji}$, that same condition now has the role of selecting internal signals to be externally excited for single module identifiability, according to the results of Theorem~\ref{theorem:SingleCutNew}. Moreover, a set of signals, which satisfies the parallel path and loop condition, can now effectively be found by constructing a (minimum) disconnecting set.

%
\section{Signal allocation for generic identifiability} \label{sec:synthesis}
If $G_{j \bar{\mathcal{W}}_j }$ is not generically identifiable in a given network model set, extra excitation signals can be allocated to achieve generic identifiability of $G_{j \bar{\mathcal{W}}_j }$. In this section an algorithm will be developed to address this synthesis problem by exploiting condition~\ref{item2} in Theorem~\ref{theorem:SingleCutNew}.

Consider a network model set with a set of initial external signals $\mathcal{X}^0_j$ that do not have unknown directed edges to $w_j$. The algorithms aim to allocate a minimum number of additional excitation signals $\mathcal{X}^a_j$ such that generic identifiability of $G_{j\bar{\mathcal{W}}_j}$ is guaranteed. In this synthesis problem, it is assumed that if $r_k$ is allocated directly at $w_j$, its corresponding transfer function $R_{jk}$ is known.

Condition \ref{item2} of Theorem~\ref{theorem:SingleCutNew} is based on the existence of a $\bar{\mathcal{X}}_j - \mathcal{W}_j \setminus \bar{\mathcal{W}}_j$ disconnecting set given $\bar{\mathcal{X}}_j$; however, the set $\bar{\mathcal{X}}_j$ depends on the allocated excitation signals and thus is not available before signal allocation. Therefore, in the synthesis approach we first find a candidate disconnecting set $\mathcal{D}$ without relying on the excitation signals, and then construct $\bar{\mathcal{X}}_j$ by allocating signals such that (i) $\mathcal{D}$ becomes a $\bar{\mathcal{X}}_j - \mathcal{W}_j \setminus \bar{\mathcal{W}}_j$ disconnecting set and (ii) $\mathcal{D} \cup \bar{\mathcal{W}}_j$ are excited by $\bar{\mathcal{X}}_j$ to satisfy \eqref{eq:TheoremCut}. A candidate set $\mathcal{D}$ can be motivated by the following corollary.

\begin{corollary} \label{lemma:VDpath2}
Assume that one of the conditions in Theorem \ref{theorem:SingleCutNew} is satisfied, then $\mathcal{D}$ is a $\bar{\mathcal{X}}_j - \mathcal{W}_j \setminus \bar{\mathcal{W}}_j$ disconnecting set if and only if it is a $\bar{\mathcal{W}}_j \cup \bar{\mathcal{X}}_j  - \mathcal{W}_j \setminus \bar{\mathcal{W}}_j$ disconnecting set subject to $\mathcal{D} \cap \bar{\mathcal{W}}_j = \emptyset$.
\end{corollary}
\begin{pf}
The ``if" part is trivial. The ``only if" part is proved by contradiction. If $\mathcal{D}$ is not a $\bar{\mathcal{W}}_j - \mathcal{W}_j \setminus \bar{\mathcal{W}}_j$ disconnecting set, i.e. there exists a path from $\bar{\mathcal{W}}_j$ to $\mathcal{W}_j \setminus \bar{\mathcal{W}}_j$ which does not intersect with $\mathcal{D}$, then there is also a path from $\mathcal{X}_j$ (or $\bar{\mathcal{X}}_j$) via $\bar{\mathcal{W}}_j$ to $\mathcal{W}_j \setminus \bar{\mathcal{W}}_j$ and the path does not intersect with $\mathcal{D}$, which contradicts that $\mathcal{D}$ is a disconnecting set. In addition, following from \eqref{eq:TheoremCut}, $\mathcal{D} \cap \bar{\mathcal{W}}_j = \emptyset$ must hold. $\hfill{} \blacksquare$
\end{pf}

The above result indicates that a $\bar{\mathcal{W}}_j  -  \mathcal{W}_j \setminus \bar{\mathcal{W}}_j$ disconnecting set $\mathcal{D}$ subject to $\mathcal{D} \cap \bar{\mathcal{W}}_j = \emptyset$ can be computed first, which is independent of the allocated external signals. Then extra excitation signals can be allocated such that signals in $\mathcal{D}\cup \bar{\mathcal{W}}_j$ are excited, and $\mathcal{D}$ becomes a disconnecting set from the allocated signals to $ \mathcal{W}_j \setminus \bar{\mathcal{W}}_j$.

As the number of required excitation signals depends on the cardinality of the disconnecting set, a minimum disconnecting set is desired. Additionally, based on Corollary~\ref{lemma:VDpath2}, a minimum disconnecting set $\mathcal{D}$ subject to $\bar{\mathcal{W}}_j \cap \mathcal{D} = \emptyset$ needs to be found. As standard graphical algorithms for computing minimum disconnecting sets do not take into account any constraint, we redefine the disconnecting set to make standard algorithms applicable.
\begin{proposition} \label{pro:fresh}
Given a model set $\mathcal{M}$ and any subset $\bar{\mathcal{X}}_{j} \subseteq \mathcal{X}_{j}$, $\mathcal{D}$ is a minimum disconnecting set from $\bar{\mathcal{W}}_j \cup \bar{\mathcal{X}}_{j}$ to $\mathcal{W}_j \setminus \bar{\mathcal{W}}_j$ subject to $\bar{\mathcal{W}}_j \cap \mathcal{D} = \emptyset$ if and only if $\mathcal{D}$ is a minimum disconnecting set from $\mathcal{N}_{\bar{\mathcal{W}}_j}^+ \cup \bar{\mathcal{X}}_{j}$ to $\mathcal{W}_j \setminus \bar{\mathcal{W}}_j$.
\end{proposition}
\begin{pf}
We will first show that for any vertex set $\mathcal{D}$ subject to $\bar{\mathcal{W}}_j \cap \mathcal{D} = \emptyset$, $\mathcal{D}$ is a disconnecting set from $\mathcal{N}_{\bar{\mathcal{W}}_j}^+ \cup \bar{\mathcal{X}}_{j}$ to $\mathcal{W}_j \setminus \bar{\mathcal{W}}_j$ if and only if it is also a disconnecting set from $\bar{\mathcal{W}}_j \cup \bar{\mathcal{X}}_j$ to $\mathcal{W}_j \setminus \bar{\mathcal{W}}_j$. The ``only if" part holds because if $\mathcal{D}$ intersects with all paths from $\mathcal{N}_{\bar{\mathcal{W}}_j}^+ \cup \bar{\mathcal{X}}_j$ to $\mathcal{W}_j \setminus \bar{\mathcal{W}}_j$, then it also intersects with the paths from $\bar{\mathcal{W}}_j \cup \bar{\mathcal{X}}_j$ to $\mathcal{W}_j \setminus \bar{\mathcal{W}}_j$. For the ``if" part, since $\bar{\mathcal{W}}_j \cap \mathcal{D} = \emptyset$ and $\mathcal{D}$ intersects with all the paths from $\bar{\mathcal{W}}_j \cup \bar{\mathcal{X}}_j$ to $\mathcal{W}_j \setminus \bar{\mathcal{W}}_j$, those paths from $\bar{\mathcal{W}}_j$ to $\mathcal{W}_j \setminus \bar{\mathcal{W}}_j$ have to intersect with $\mathcal{D}$ at their internal vertices or the ending vertices. Since the first internal vertices of the paths belong to set $\mathcal{N}_{\bar{\mathcal{W}}_j}^+$, then $\mathcal{D}$ is also a disconnecting set from $\mathcal{N}_{\bar{\mathcal{W}}_j}^+ \cup \bar{\mathcal{X}}_j$ to $\mathcal{W}_j \setminus \bar{\mathcal{W}}_j$. \\
Having the above result, the proposition is proved by showing that a minimum disconnecting set $\mathcal{D}$ from $\mathcal{N}_{\bar{\mathcal{W}}_j}^+ \cup \bar{\mathcal{X}}_j$ to $\mathcal{W}_j \setminus \{w_i\}$ does not contain $\bar{\mathcal{W}}_j$, because if it does, it remains a disconnecting set after $\bar{\mathcal{W}}_j$ is excluded, which contradicts the minimality of $\mathcal{D}$. $\hfill{} \blacksquare$
\end{pf}

The above result shows that a minimum disconnecting set $\mathcal{D}$ from $\mathcal{N}_{\bar{\mathcal{W}}_j}^+$ to $\mathcal{W}_j \setminus \bar{\mathcal{W}}_j$ can be computed for the synthesis problem, which now is an unconstrained problem and thus can be solved by standard graphic algorithms, e.g. the Ford-Fulkerson algorithm \citep{schrijver2003combinatorial}. Having established a choice of $\mathcal{D}$, the following synthesis result can be derived from condition~\ref{item2} in Theorem~\ref{theorem:SingleCutNew}.

\begin{corollary} \label{corro:AlloSingleNew1}
Consider a network model set $\mathcal{M}$ that satisfies Assumptions~\ref{ass:feedthrough}, \ref{ass:indePara}, \ref{ass5}. Given any minimum disconnecting set $\mathcal{D}$ from $\mathcal{N}^+_{\bar{\mathcal{W}}_j}$ to $\mathcal{W}_j \setminus \bar{\mathcal{W}}_j$, assigning distinct excitation signals to every vertex in $\mathcal{D} \cup \bar{\mathcal{W}}_j$ leads to generic identifiability of $G_{j\bar{\mathcal{W}}_j}$ in $\mathcal{M}$ from $(w,r)$.
\end{corollary}
\begin{pf}
Let $\mathcal{X}^a_j$ denote the set of allocated signals, and $\mathcal{X}^a_j \subseteq \mathcal{X}_j$ holds in the obtained model set after allocation, i.e. $\mathcal{X}^a_j$ has no unknown directed edge to $w_j$. As these signals are allocated directly at $\mathcal{D}\cup \bar{\mathcal{W}}_j$, $\mathcal{D}$ is a $\mathcal{N}_{\bar{\mathcal{W}}_j}^+ \cup \mathcal{X}^a_j - \mathcal{W}_j \setminus \{w_i\} $ and thus a $\bar{\mathcal{W}}_j \cup \mathcal{X}^a_j - \mathcal{W}_j \setminus \{w_i\} $ disconnecting set, based on Proposition~\ref{pro:fresh}. In addition, \eqref{eq:TheoremCut} holds with $\bar{\mathcal{X}}_j = \mathcal{X}^a_j$. Thus, Theorem~\ref{theorem:SingleCutNew}\ref{item2} is satisfied with $\bar{\mathcal{X}}_j = \mathcal{X}^a_j$ and the given $\mathcal{D}$. $\hfill{} \blacksquare$
\end{pf}

The above result provides the basis for a synthesis approach that first computes a disconnecting set $\mathcal{D}$ and then allocates external signals to directly excite $\mathcal{D}\cup \bar{\mathcal{W}}_j$. This approach is valid even if a non-minimum disconnecting set is considered. However, it does not
consider the initially present signals $\mathcal{X}^0_j$, and thus may allocate redundant signals. Additionally, excitation signals do not necessarily have to be directly allocated at the vertices in $\mathcal{D} \cup \bar{\mathcal{W}}_j$, but could also reach those vertices through vertex disjoint paths.
%
To make use of $\mathcal{X}^0_j$ and to explore the freedom to allocate the signals, a more comprehensive method is introduced in Algorithm~1.

Given a network model set with the target modules $G_{j\bar{\mathcal{W}}_j}$ and the pre-existing external signals $\mathcal{X}_j^0$, Algorithm~1 computes a minimum disconnecting set $\mathcal{D}$ from $\mathcal{N}^+_{\bar{\mathcal{W}}_j} \cup \mathcal{X}^0_j$ to $\mathcal{W}_j \setminus \bar{\mathcal{W}}_j$ first and then removes the vertices in $\mathcal{D}\cup \bar{\mathcal{W}}_j$ that are already excited by $\mathcal{X}^0_j$ via vertex disjoint paths. Then the algorithm allocates additional excitation signals to excite the remaining vertices in $\mathcal{D}\cup \bar{\mathcal{W}}_j$ through vertex disjoint paths. The validity of Algorithm~1 is shown in the following result.
\begin{figure*}[h]
\begin{minipage}{0.32\textwidth}
\centering
\includegraphics[scale=0.3]{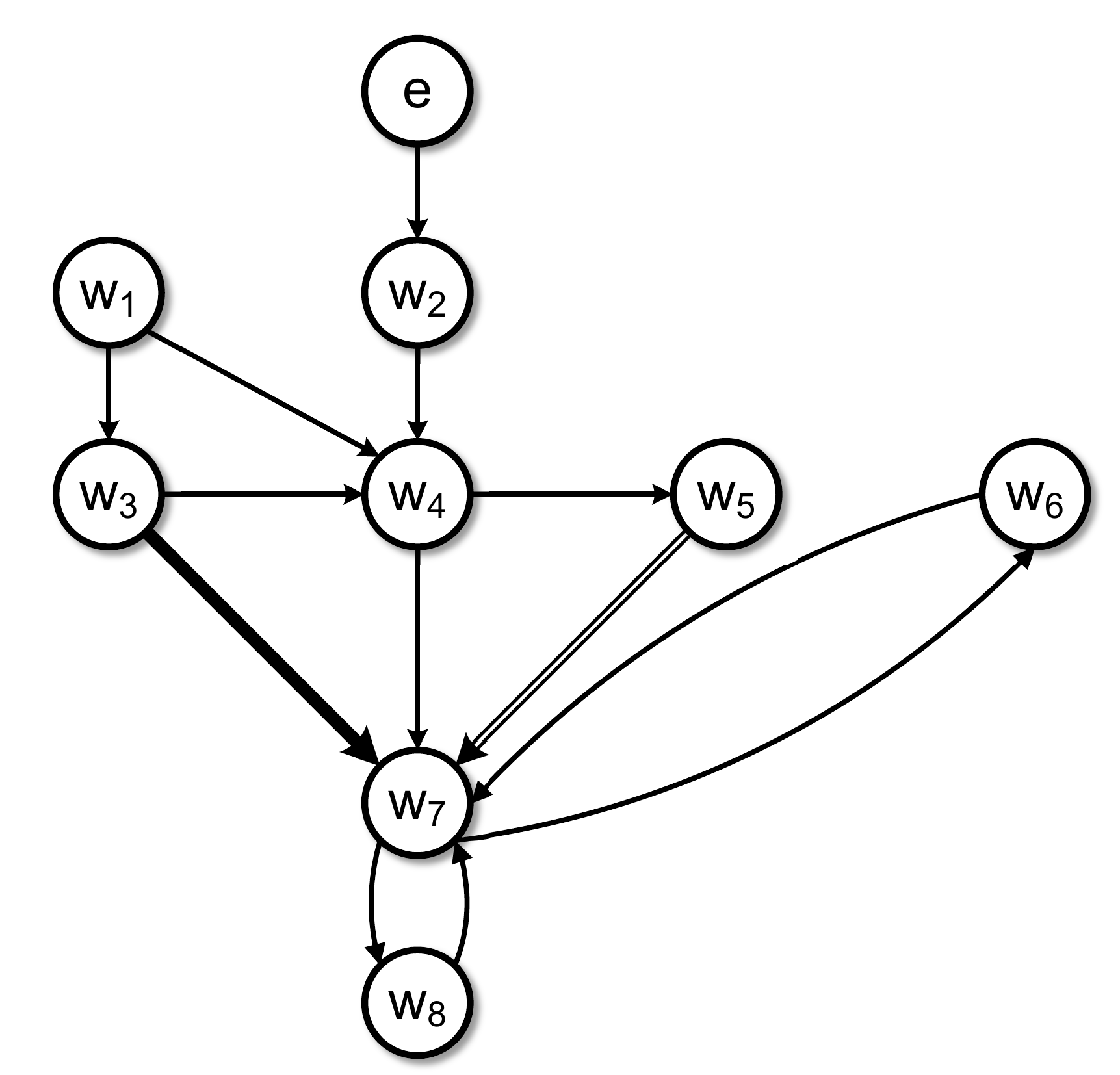}
\\ (a)
\end{minipage}
\begin{minipage}{0.32\textwidth}
\centering
\includegraphics[scale=0.3]{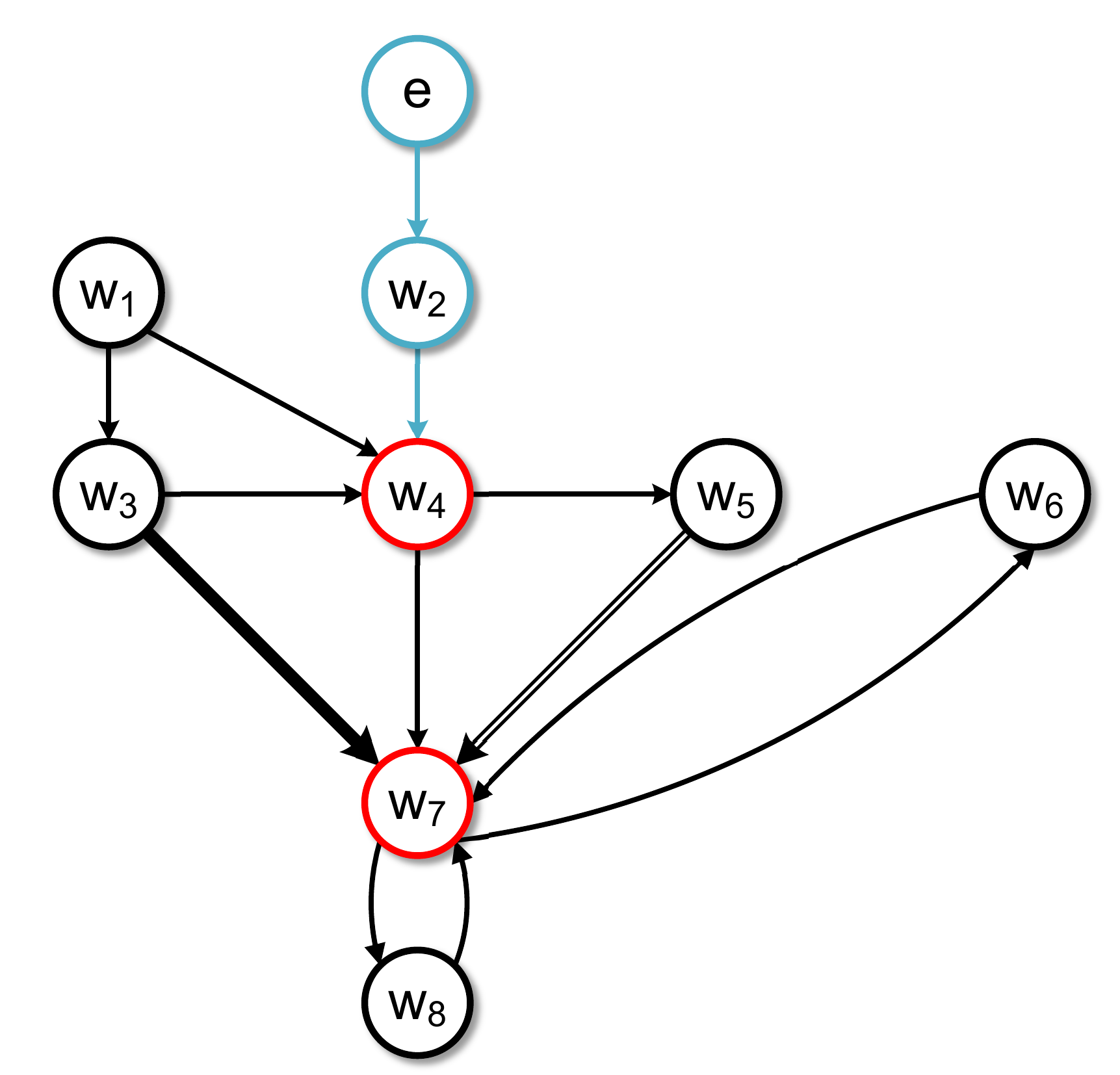}
\\ (b)
\end{minipage}
\begin{minipage}{0.32\textwidth}
\centering
\includegraphics[scale=0.3]{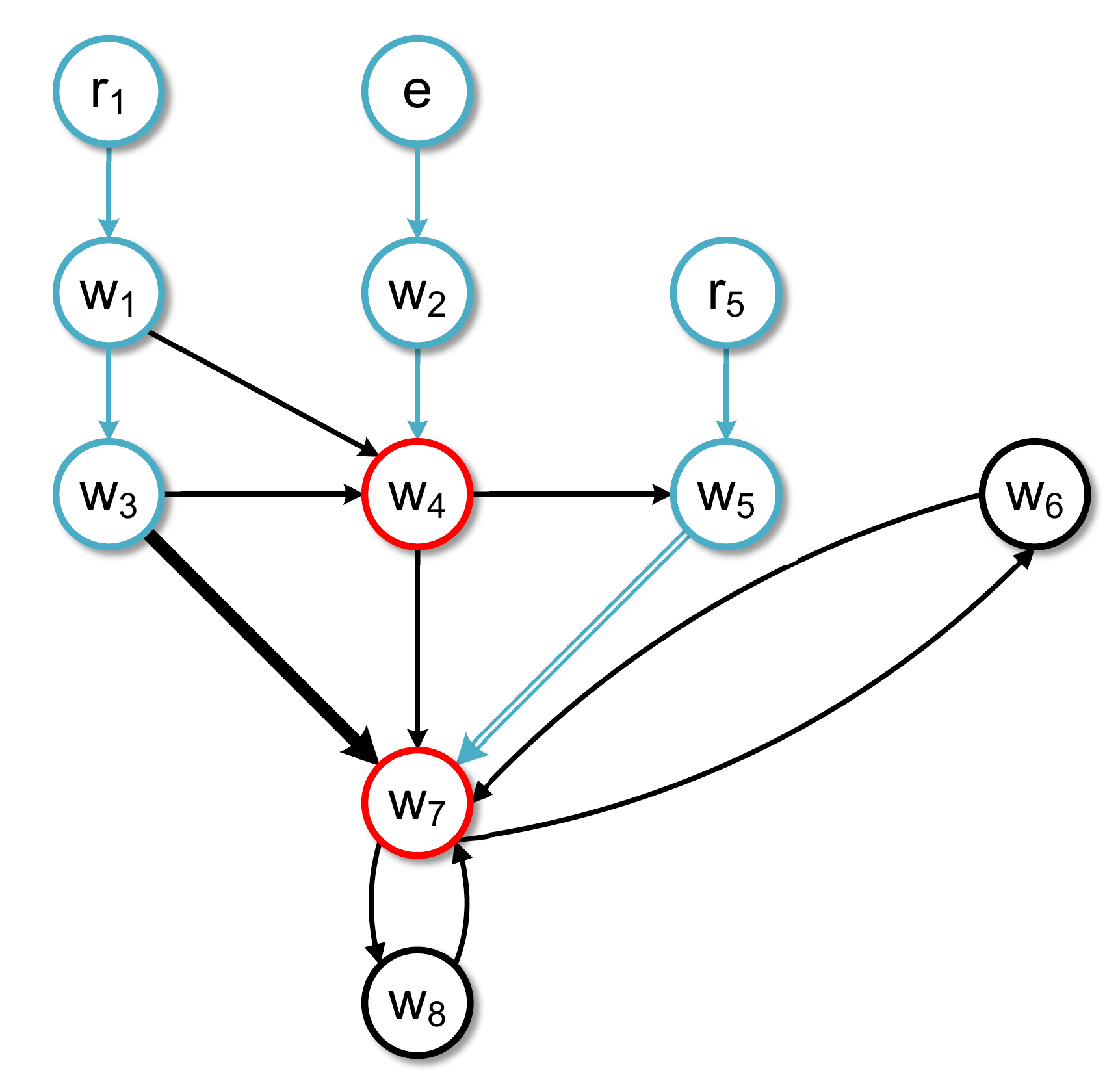}
\\ (c)
\end{minipage}
\caption{An example of allocating signals for generic identifiability of $G_{73}$ (thick line) using Algorithm~\ref{alg:AssignSingeNew} with a known module $G^0_{75}$ (double-line edge). Starting from the network model in (a), a disconnecting set (red vertices) is computed in (b). Since there already exists an external signal $e$, which has a path to $w_4$ that belongs to the disconnecting set, we only need to add $r_1$ and $r_5$ as in (c), which achieves generic identifiability of $G_{73}$.}
\label{fig:OneModuleCutNew}
\end{figure*}

\begin{theorem}
Given a network model set $\mathcal{M}$ that satisfies Assumptions~\ref{ass:feedthrough}, \ref{ass:indePara}, \ref{ass5}. In the returned model set of Algorithm~1, $G_{j\bar{\mathcal{W}}_j}$ is generically identifiable from $(w,r)$.
\end{theorem}
\begin{pf}
From step $1$ to $2$ in Algorithm~\ref{alg:AssignSingeNew}, by construction, if $|\mathcal{P}|= |\mathcal{D} \cup \bar{\mathcal{W}}_j|$, (\ref{eq:TheoremCut}) holds with $\bar{\mathcal{X}}_j = \mathcal{X}_j^0 $, and thus the modules are generically identifiable in the original model set $\mathcal{M}$. When $|\mathcal{P}| < |\mathcal{D} \cup \bar{\mathcal{W}}_j|$, based on Theorem~\ref{theorem:SingleCutNew}, we need to allocate extra $|\mathcal{D} \cup \bar{\mathcal{W}}_j|- |\mathcal{P}| $ signals, such that: (i) there are $|\mathcal{D} \cup \bar{\mathcal{W}}_j|- |\mathcal{P}| $ vertex disjoint paths from these signals to $(\mathcal{D} \cup \bar{\mathcal{W}}_j) \setminus \bar{\mathcal{D}}$, and these paths are also vertex disjoint with $\mathcal{P}$; (ii) $\mathcal{D}$ remains a disconnecting set from the added signals to $\mathcal{W}_j \setminus \bar{\mathcal{W}}_j$. Then we can find in the algorithm, steps $5$ guarantees (ii), and steps $3$, $6$, $7$ and $8$ guarantee (i). This concludes the proof.
$\hfill{} \blacksquare$
\end{pf}

\begin{algorithm}[t]
\caption{Signal allocation for local modules}
\begin{flushleft}
        \textbf{INPUT:} A model set $\mathcal{M}$ with graph $\mathcal{G}$,  the target modules $G_{j \bar{\mathcal{W}}_j}$, and a set of initial external signals $\mathcal{X}^0_j$; \\
        \textbf{OUTPUT:} A new model set $\mathcal{M}_{out}$ with its graph $\mathcal{G}_{out}$
\end{flushleft}
\begin{algorithmic}[1]
\STATE Compute a minimum disconnecting set $\mathcal{D}$ from $\mathcal{N}^+_{\bar{\mathcal{W}}_j} \cup \mathcal{X}^0_j$ to $\mathcal{W}_j \setminus \bar{\mathcal{W}}_j$;
\STATE Based on Lemma~\ref{lemma:VDpath0}, compute a set $\mathcal{P}$ that contains the maximum number of vertex disjoint paths from $\mathcal{X}^0_j$ to $\mathcal{D} \cup \bar{\mathcal{W}}_j$, while the paths are internally vertex disjoint with $\mathcal{D} \cup \bar{\mathcal{W}}_j$;

\STATE Let $\bar{\mathcal{D}} \subseteq \mathcal{D} \cup \bar{\mathcal{W}}_j$ denote all the ending vertices of the paths in $\mathcal{P}$;


\IF{$|\mathcal{P}| < |\mathcal{D} \cup \bar{\mathcal{W}}_j|$}
\STATE Find the largest set $\bar{\mathcal{W}} \subseteq \mathcal{W}$ such that $\mathcal{D}$ is a disconnecting set from $\bar{\mathcal{W}}$ to $\mathcal{W}_j \setminus \bar{\mathcal{W}}_j$;

\STATE Build a subgraph $\bar{\mathcal{G}} \subseteq \mathcal{G}$ by removing all vertices and edges of the paths in $\mathcal{P}$;

\STATE  Find a set $\mathcal{W}_{exp} \subseteq \bar{\mathcal{W}}$ such that in $\bar{\mathcal{G}}$, there are $|\mathcal{D} \cup \bar{\mathcal{W}}_j| -|\mathcal{P}|$ vertex disjoint paths from $\mathcal{W}_{exp}$ to $(\mathcal{D} \cup\bar{\mathcal{W}}_j) \setminus \bar{\mathcal{D}}$ ;

\STATE In $\mathcal{G}$, assign distinct excitation signals to every vertex in $\mathcal{W}_{exp}$, which leads to a new model set $\mathcal{M}_{out}$ with a new graph $\mathcal{G}_{out}$ and a new $R$ matrix;
\STATE  Return $\mathcal{M}_{out}$ with the graph $\mathcal{G}_{out}$;

\ELSE
\STATE $\mathcal{M}_{out} \gets \mathcal{M} $ and $\mathcal{G}_{out} \gets \mathcal{G}$;
\STATE Return $\mathcal{M}_{out}$ with the graph $\mathcal{G}_{out}$.
\ENDIF
\end{algorithmic}
\label{alg:AssignSingeNew}
\end{algorithm}
\begin{remark}
In Algorithm~\ref{alg:AssignSingeNew}, the signals do not need to be allocated directly at the vertices in $\mathcal{D} \cup \bar{\mathcal{W}}_j$. It suffices when they reach these vertices through vertex disjoint paths from excitation locations elsewhere in the network.
\end{remark}

Algorithm~\ref{alg:AssignSingeNew} and the synthesis approach in Corollary~\ref{corro:AlloSingleNew1} guarantee a minimum number of allocated signals when $\mathcal{X}^0_j = \emptyset$, as a minimum disconnecting set is used. While a minimum number of allocated signals can not be guaranteed when $\mathcal{X}^0_j \not= \emptyset$, the following bound on the number of additionally required signals can be derived.

\begin{corollary}
Given a network model set $\mathcal{M}$ that satisfies Assumptions~\ref{ass:feedthrough}, \ref{ass:indePara}, \ref{ass5}. Let $\mathcal{X}_j^0$ denote the set of initial external signals that have no unknown directed edge to $w_j$. Let $\mathcal{D}$ be a minimum disconnecting set from $\mathcal{N}_{\bar{\mathcal{W}}_j}^+ \cup \mathcal{X}^0_j$ to $\mathcal{W}_j \setminus \bar{\mathcal{W}}_j$, and let $c$ denote the number of excitation signals available for allocation. Then the number of additional excitation signals $c$ is sufficient to make $G_{j\bar{\mathcal{W}}_j}$ generically identifiable if $
c \geqslant |\mathcal{D} \cup \bar{\mathcal{W}}_j| - b_{\mathcal{X}^0_j \to \mathcal{D} \cup \bar{\mathcal{W}}_j}.
$
\end{corollary}

Algorithm~1 is illustrated in the following example.
\begin{example}
In the network model in Fig.~\ref{fig:OneModuleCutNew}(a), the problem is to allocate excitation signals such that $G_{73}$ becomes generically identifiable. $\mathcal{X}^0_{7} = \{e\}$ is the only external signal that is initially present. Firstly, a disconnecting set from $\mathcal{X}^0_{7} \cup \{w_3\}= \{w_3,e\}$ to the other in-neighbors of $w_7$ through unknown edges, i.e. $\mathcal{W}_7 \setminus \{w_3\} = \{w_4,w_6,w_8\}$, is constructed as $\mathcal{D}= \{w_4,w_7\}$, indicated in Fig.~\ref{fig:OneModuleCutNew}(b). Based on Theorem~\ref{theorem:SingleCutNew}, generic identifiability of $G_{73}$ requires three vertex disjoint paths from external signals to $\mathcal{D} \cup \{w_3\} =\{w_3,w_4,w_7\}$, while $\mathcal{D}$ remains a disconnecting set from the external signals to $\mathcal{W}_7 \setminus \{w_3\}$. Following step $2$ in Algorithm~\ref{alg:AssignSingeNew}, we find a path $e \rightarrow w_4$ from $\mathcal{X}^0_{7}$ to $\mathcal{D} \cup \{w_3\}$ (colored blue in Fig.~\ref{fig:OneModuleCutNew}(b)). Thus we only need to allocate extra excitation signals from which there are two vertex disjoint paths to $\{w_3,w_7\}$, and the two paths should be vertex disjoint with path $e \rightarrow w_4$. As in step $5$, the set of potential locations to allocate excitation signals is given by $\bar{\mathcal{W}} = \{ w_1,w_2,w_3,w_4,w_5,w_7\}$, which satisfies the requirement that $\mathcal{D}$ remains a disconnecting set from $\bar{\mathcal{W}}$ to $\mathcal{W}_7 \setminus \{w_3\}$. Then we choose $\mathcal{W}_{exp} = \{w_1,w_5\} \subseteq \bar{\mathcal{W}} $ to be excited by $r_1$ and $r_5$, as shown in Fig.~\ref{fig:OneModuleCutNew}(c). According to Theorem~\ref{theorem:SingleCutNew}, $G_{73}$ is indeed generically identifiable in Fig.~\ref{fig:OneModuleCutNew}(c): There exists a disconnecting set $\mathcal{D} = \{w_4,w_7\}$ from $\mathcal{X}_7 = \{r_1,r_5,e\}$ to $\mathcal{W}^-_7 \setminus \{w_3\} = \{w_4,w_6,w_8\}$ such that $b_{\mathcal{X}_7 \to \mathcal{D}\cup \{w_3\} } = 3$, i.e. there are maximally three vertex disjoint paths from $\mathcal{X}_7$ to $\mathcal{D}\cup\{w_3\}$, as indicated by the blue paths in Fig.~\ref{fig:OneModuleCutNew}(c).
\label{example3}
\end{example}

When generic identifiability of $\{G_{j\bar{\mathcal{W}}_j}| j \in \{1,\cdots,L\}\}$ is of interest, Algorithm~1 can be applied recursively for each $j$. If generic identifiability of the full network is concerned, a compact graphical approach for signal allocation is presented in \citep{chengallocation,2019arXivChen}, where effective use is made of the interdependence of the $L$ different subproblems.

\section{Indirect identification method} \label{sec:indirect}
In the previous sections, graphical tools have been developed to verify identifiability and achieve identifiability via signal allocation. However, the identifiability results do not point directly to an identification method to actually estimate the modules of interest from data. In this section we will show that, when the identifiability conditions in Theorem~\ref{theorem:SingleCutNew} are satisfied, in a particular situation the identifiability results directly point to an identification method for estimating $G_{j\bar{\mathcal{W}_j}}$. The particular situation is that the identifiability conditions are satisfied on the basis of a set of external signals $\bar{\mathcal{X}}_j$ that contains $r$ signals only. We will first show how the target modules can be reconstructed on the basis of transfer functions between measured signals only, by combining the identifiability result and Theorem \ref{the:Decompo2}.
For simplicity, we consider the setting where all the known transfer operators in the network are fixed zeros.
\begin{corollary} \label{cor5}
Consider a model set $\mathcal{M}$ that has all the known transfer operators being zeros and satisfies Assumptions~\ref{ass:feedthrough}, \ref{ass:indePara}. If condition~\ref{item2} of Theorem~\ref{theorem:SingleCutNew} is satisfied with $\bar{\mathcal{X}}_j$ having no directed edge to $w_j$, then it holds that
\begin{equation} \label{eq:gfac}
G_{j\bar{\mathcal{W}}_j}(q,\theta) = T_{j\bar{\mathcal{X}}_j}(q,\theta)  \begin{bmatrix} T_{\bar{\mathcal{W}}_j\bar{\mathcal{X}}_j}(q,\theta) \\ T_{\mathcal{D}\bar{\mathcal{X}}_j}(q,\theta) \end{bmatrix}^{\dagger}
\begin{bmatrix} I_{|\bar{\mathcal{W}}_j|} \\ 0 \end{bmatrix},
\end{equation}
where $(\cdot)^{\dagger}$ denotes the right inverse and exists for almost all $\theta $, and $ I_{|\bar{\mathcal{W}}_j|}$ denotes an identity matrix of size $|\bar{\mathcal{W}}_j|$.
\end{corollary}
\begin{pf}
When all the known transfer operators in $\mathcal{M}$ are zeros, Assumption~\ref{ass5} is not needed, and $\mathcal{W}_j$ contains all the internal signals having directed edges to $w_j$. Then the permutation of the $j$th row in $(I-G)T_{\mathcal{W}\mathcal{X}} = X$ leads to
\begin{equation}
\begin{bmatrix}
- G_{j\bar{\mathcal{W}}_j} & -G_{j(\mathcal{W}_j \setminus \bar{\mathcal{W}}_j)}& 1
\end{bmatrix} \begin{bmatrix}
T_{\bar{\mathcal{W}}_j \bar{\mathcal{X}}_j} \\
T_{(\mathcal{W}_j \setminus \bar{\mathcal{W}}_j)\bar{\mathcal{X}}_j} \\
T_{j \bar{\mathcal{X}}_j}
\end{bmatrix} = X_{j\bar{\mathcal{X}}_j}, \label{eq:ProofFinalCoro}
\end{equation}
where $X_{j\bar{\mathcal{X}}_j}=0$ as $\bar{\mathcal{X}}_j$ has no edge to $w_j$. If condition \ref{item2} of Theorem~\ref{theorem:SingleCutNew} holds, there exists a $K$ so that $T_{(\mathcal{W}_{j} \setminus \bar{\mathcal{W}}_j)\bar{\mathcal{X}}_j} = K T_{\mathcal{D}\bar{\mathcal{X}}_j }$ according to Theorem \ref{the:Decompo2}. Combing the above decomposition and \eqref{eq:ProofFinalCoro} leads to \eqref{eq:gfac}, where the right inverse exists generically due to \eqref{eq:TheoremCut} and \eqref{eq:vdp}.
$\hfill{} \blacksquare$
\end{pf}

The expression (\ref{eq:gfac}) shows an immediate opportunity to estimate $G_{j\bar{\mathcal{W}}_j}$ based on a selected set of measured internal signals. When $\bar{\mathcal{X}}_j$ contains only measured excitation signals $r$, the transfer functions on the right hand side of (\ref{eq:gfac}) can all be estimated consistently through a standard identification method, based on the measured signals in $\bar{\mathcal{X}}_j \cup \bar{\mathcal{W}}_j\cup \mathcal{D} \cup \{w_j\}$.
\\
The above method is a generalization of the ``classical'' indirect method of identification, as developed for closed-loop systems \cite{ljung1987system} and dynamic networks \cite{gevers2018practical}, where all the in-neighbors of $w_j$ are measured and directly excited for identifying the target modules. The method based on Corollary~\ref{cor5} generalizes this situation by allowing to measure and excite the internal signals determined by disconnecting sets. This generalization can lead to an experimental setup with fewer actuators and sensors for identifying the target modules than required by the classical method.

\begin{remark}
When $\bar{\mathcal{X}}_j$ only contains $r$ signals, it is shown in this section that $G_{j\bar{\mathcal{W}}_j}$ can be estimated consistently based on the signals in $\mathcal{D} \cup \bar{\mathcal{W}}_j\cup \{w_j\}$. It is then straightforward that under these conditions $G_{j\bar{\mathcal{W}}_j}$ is generically identifiable in an appropriately chosen model set, without requiring to measure all the internal signals. This establishes a particular generic identifiability result with unmeasured internal signals, as also addressed in \citep{bazanella2019network}.
\end{remark}
If the set $\bar{\mathcal{X}}_j$ contains both $r$ and $e$-signals, then the reasoning underlying the indirect approach fails, and one can apply a direct method for consistently identifying the target module, see \cite{ramaswamy2019local}.

\section{Duality}
It is not hard to extend the results of this work from the full measurement setting to the full excitation setting in \cite{bazanella2017identifiability} \citep{hendrickx2018identifiability}, where all internal signals are excited with a subset unmeasured, and only $r$ signals are used as excitation sources for identifiability. In this case, the goal is to ensure the uniqueness of a subset of out-going edges of $w_i$, given the mapping from the $r$ signals to the measured internal signals. Then the path-based condition as in Theorem~V.I of \citep{hendrickx2018identifiability} can be obtained, and a disconnecting-set-based result can be obtained similarly.\\
However, exploiting the noise signals as excitation sources for identifiability, as done in Proposition~\ref{eq:finalDefn}, is not trivial when there are unmeasured internal signals. This requires to handle the noise information in a submatrix of the power spectrum $\Phi_{\bar{v}}$ of the disturbance, instead of the full matrix $\Phi_{\bar{v}}$ as considered in this work. This more complex situation is further addressed in the follow up work \cite{Shi&Cheng&VandenHof:20}.

%
%
%

\section{Conclusion} \label{sec:conclu}
In this work, generic identifiability of a subnetwork, i.e. a subset of modules, in linear dynamic networks has been investigated in the setting where all internal signals are measured. Algebraic conditions are obtained for both generic and global identifiability. Then a path-based condition for generic identifiability is obtained by extending the results in \citep{hendrickx2018identifiability}, in order to handle the model-set type of identifiability, while including disturbance inputs in the network and allowing network modules to be fixed/known. This path-based condition is then equivalently reformulated into a novel graphical condition based on disconnecting sets. With this new condition, the synthesis problem is tackled, which aims to allocate the minimum number of excitation signals to achieve generic identifiability of the modules. In addition, the new condition allows us to make a further step towards a new indirect identification method for estimating the subnetwork. The next step in this research is the handling of the more complex situation of partial measurement and partial excitation. This topic is addressed in the follow up work \cite{Shi&Cheng&VandenHof:20}. The Matlab software for conducting identifiability analysis based on Theorem \ref{theorem:PathCondi} and synthesis according to Algorithm~\ref{alg:AssignSingeNew} can be found in \citep{ShiCodeFull}.


\section*{Appendix}

\subsection*{Proof of Theorem~\ref{theorem:Goodrank}}
The following lemma is instrumental.
\begin{lemma} \label{lemma:nece1}
For a model set $\mathcal{M}$ with a graph $\mathcal{G}$ that satisfies Assumption~\ref{ass:feedthrough}, consider the corresponding set $\mathbb{G}^\star$ as in Definition~\ref{def:strucModelSet}. If $M(\theta_0) \in \mathcal{M}$ does not satisfy \eqref{eq:rankCondi}, then for any positive real number $r$, there exists another model $\bar{M}=(\bar{G}(q),R(q,\theta_0),H(q,\theta_0),\Lambda(\theta_0))$, such that (i) $\bar{G}(q)$ differs from $G(q,\theta_0)$ only in $\bar{G}_{j\bar{\mathcal{W}_j}}(q)$ (ii) $\bar{G}(q) \in \mathbb{G}^\star$; (iii) $0< ||G(q,\theta_0)-\bar{G}(q)||_{\infty} < r $; (iv) $T_{\mathcal{W}\mathcal{X}}(q,\theta_0) = \bar{T}_{\mathcal{W}\mathcal{X}}(q)$.
\end{lemma}
\begin{pf}
Recall the set $\mathcal{X}_j$. Consider $j$th row of $(I-G)T_{\mathcal{W} \mathcal{X}} = X$ and its columns corresponding to signals in $\mathcal{X}_j$, and after permutation, it leads to$$
\begin{bmatrix}
-G_{j \bar{\mathcal{W}}_j}& -G_{j (\mathcal{W}_j \setminus \bar{\mathcal{W}}_j}) & -G_{j \widetilde{W}_j} & 1 & 0
\end{bmatrix}
T_{\star \mathcal{X}_j} = X_j,
$$ where $X_j$ is a known vector by the definition of $\mathcal{X}_j$, $G_{j \widetilde{W}_j}$ contains all known non-zero entries in the $j$th row of $(I-G)$. Thus, the above equation leads to
\begin{equation}
\begin{bmatrix}
-G_{j \bar{\mathcal{W}}_j } & -G_{j (\mathcal{W}_j \setminus \bar{\mathcal{W}}_j)}
\end{bmatrix}
\begin{bmatrix}
T_{\bar{\mathcal{W}}_j \mathcal{X}_j}   \\
T_{(\mathcal{W}_j \setminus \bar{\mathcal{W}}_j) \mathcal{X}_j}
\end{bmatrix} = P, \label{eq:theoremRankProof1}
\end{equation}
where $P = X_j - T_{j\mathcal{X}_j}+ G_{j \widetilde{W}_j} T_{\widetilde{W}_j\mathcal{X}_j}$ and $P$ is available. Then based on Definition~\ref{eq:finalDefn}, identifiability concerns if a unique vector $G_{j \bar{\mathcal{W}}_j }$ can be obtained given $T_{\mathcal{W} \mathcal{X}}$ and $P$, for (almost) all models in $\theta \in \Theta$. Denote $G(q,\theta_0)$ by $G_0$ for simplicity, and conditions \eqref{eq:6a} and \eqref{eq:6b} will be discussed separately.

If condition~\eqref{eq:6a} is not satisfied, i.e., $T_{\bar{\mathcal{W}}_j \mathcal{X}_j}$ formulated from $M(\theta_0)$ is not full rank, then according to \eqref{eq:theoremRankProof1}, there exists a nonzero transfer vector $Q$ such that $QT_{\bar{\mathcal{W}}_j \mathcal{X}_j}=0$ and thus
\begin{equation}
 \begin{bmatrix}
-(G_{j \bar{\mathcal{W}}_j }+FQ) & -G_{j (\mathcal{W}_j \setminus \bar{\mathcal{W}}_j)}
\end{bmatrix}
\begin{bmatrix}
T_{\bar{\mathcal{W}}_j \mathcal{X}_j}   \\
T_{(\mathcal{W}_j \setminus \bar{\mathcal{W}}_j) \mathcal{X}_j}
\end{bmatrix} = P, \label{eq:theoremRankProof2}
\end{equation}
where $F$ is any non-zero scalar transfer operator. Consider now a new network model $\bar{M}$ with $\bar{G}$, which is obtained by replacing only $G_{j \bar{\mathcal{W}}_j }$ in $G_0$ by $G_{j \bar{\mathcal{W}}_j }+FQ$. Our goal is then to show that there exists an $F$ and consequently a model $\bar{M}$, such that $\bar{M}$ satisfies conditions~$(ii)$, $(iii)$; while by construction, $\bar{M}$ already satisfies $(i)$ and $(iv)$ for any non-zero $F$.

To show that there exists an $F$ such that $\bar{M}$ satisfies condition~$(ii)$, and according to Definition~\ref{def:strucModelSet}, we need to find an $F$ such that: $(iia)$ $\bar{G}$ satisfies Assumptions~\ref{ass1} (a), (b), (c) and Assumption~\ref{ass:feedthrough}; $(iib)$ $\bar{G}$ has the feedthrough structure as $G(q,\theta)$; and $(iic)$ $\bar{G}$ has the same fixed entries as $G_0$. To this end, let $F=a F_1$ where $a$ is an arbitrarily small positive real number and $F_1$ satisfies the following conditions: $F_1$ is stable and has zeros equal all unstable poles of $Q$; it has a delay of sufficiently high order such that $F_1 Q$ is strictly proper. Then with $F=aF_1$, it is straightforward that $\bar{G}$ satisfies Assumptions~\ref{ass1}(a), (b), Assumption~\ref{ass:feedthrough} in $(iia)$, and the conditions in $(iib)$ and $(iic)$. Finally, $\bar{G}$ can be shown to satisfy the Assumption~\ref{ass1}(c) in $(iia)$ with  $F=a F_1$ as follows. As $(I-G_0)^{-1}$ is stable, $1 / \det(I-G_0)$ is also stable. Based on the Laplace formula, it holds that
\begin{equation}
\frac{1}{\det(I-G_0) } = \frac{1}{1+[ \sum_{i=1}^L (-1)^{i+j}g_{ji} M_{ji}-1]}
\end{equation}
where $M_{ji}$ is the $(j,i)$ minor of $I-G_0$ and $g_{ji}$ is the $(j,i)$ entry of $I-G_0$. Define that $L\triangleq  \sum_{i=1}^L (-1)^{i+j}g_{ji} M_{ji}-1$, and based on the Nyquist stability theorem, the Nyquist plot of $L(jw)$ for $w \in \mathbb{R}$ in the complex domain does not encircle point $(-1,0)$, because $I-G_0$ is stable and thus $L$ is stable. Now consider $\bar{G}$, which differs from $G_0$ only in the $j$th row. It holds that
$
\det(I-\bar{G}) = \sum_{i=1}^L (-1)^{i+j}(g_{ji} +a[F_1\bar{G}]_{i})M_{ji},
$
where $[F_1\bar{G}]_{i}$ denotes the $i$th entry of vector $F_1\bar{G}$ and $[F_1\bar{G}]_{i}=0$ for some $i$. Similarly,
\begin{equation}
\frac{1}{\det(I-\bar{G}) } = \frac{1}{1+\bar{L}(z)}, \label{eq:gbar}
\end{equation}
where $\bar{L} \triangleq  L+ a\sum_{i=1}^L (-1)^{i+j} [F_1\bar{G}]_{i} M_{ji}$. As the Nyquist plot of $L(jw)$ does not encircle $(-1,0)$ and $a$ is arbitrarily small, the real part and the imaginary part of $[\bar{L}(jw) - L(jw)]$ is arbitrarily small for all $w$, and thus the Nyquist plot of $\bar{L}(jw)$ also does not encircle $(-1,0)$. This means that there exists $F=aF_1$ such that $\bar{M}$ satisfies Assumption~\ref{ass1}(c) in $(iia)$, since $1 / \det(I-\bar{G})$ and consequently $(I-\bar{G})^{-1}$ are stable because $\bar{G}$ is stable. This concludes that there exists $F=a F_1$ leading to a network model $\bar{M}$, whose matrix $\bar{G}$ satisfies condition~$(ii)$. Finally, for condition~$(iii)$, with $F=a F_1$, $ ||\bar{G}-G_0||_\infty = a|| \Delta||_\infty$, where $\Delta$ is a matrix contains the vector $F_1 Q$ and has all the other entries as zeros. As $a$ is arbitrarily small, $||\bar{G}-G_0||_\infty$ can also be made arbitrarily small, which concludes the proof for the case where condition~\eqref{eq:6a} is not satisfied.

Secondly, assuming that condition~\eqref{eq:6b} is not satisfied, then there exist a vector in the row space of $T_{\bar{\mathcal{W}}_j \mathcal{X}_j}$ which is linearly dependent on the row space of $T_{(\mathcal{W}_j \setminus \bar{\mathcal{W}}_j) \mathcal{X}_j}$. Equivalently, this means that there exists two non-zero vectors $Q_1$ and $Q_2$, such that $Q_1T_{\bar{\mathcal{W}}_j \mathcal{X}_j} + Q_2 T_{\bar{\mathcal{W}}_j \mathcal{X}_j}=0 $ and thus $P$ equals
$$
\begin{bmatrix}
-(G_{j \bar{\mathcal{W}}_j }+FQ_1) & -(G_{j (\mathcal{W}_j \setminus \bar{\mathcal{W}}_j)}+FQ_2)
\end{bmatrix} \begin{bmatrix}
T_{\bar{\mathcal{W}}_j \mathcal{X}_j}   \\
T_{(\mathcal{W}_j \setminus \bar{\mathcal{W}}_j) \mathcal{X}_j}
\end{bmatrix}$$
where $F$ is any non-zero scalar transfer operator. Now, let $\bar{G}$ be a new matrix obtained by replacing $G_{j \bar{\mathcal{W}}_j }$ and $G_{j (\mathcal{W}_j \setminus \bar{\mathcal{W}}_j)}$ in $G_0$ with $G_{j \bar{\mathcal{W}}_j } + FQ_1$ and $G_{j (\mathcal{W}_j \setminus \bar{\mathcal{W}}_j)}+FQ_2$ respectively. Then similarly, there always exists a $F$ such that $\bar{G} \in \mathbb{G}^\star$ and $||\bar{G}-G_0||_\infty$ is arbitrary small.$\hfill{} \blacksquare$
\end{pf}

The above lemma indicates that if $T_{\mathcal{W}\mathcal{X}}$ does not satisfy \eqref{eq:rankCondi}, it leads to two different matrices $\bar{G}$ and $G$ that contain different modules in $G_{j\bar{\mathcal{W}}_j}$ and are arbitrary close in the metric space $(\mathbb{G}^\star,d)$, where $d$ is a metric induced by $H_\infty$ norm. To prove the necessity of \eqref{eq:rankCondi} by contradiction, we need to show that $\bar{G}$ is in the model set under Assumptions~\ref{ass:indePara} and \ref{ass:openset}. The proof proceeds as follows.
\begin{pf}
For the ``only if" part, assume that $M(\theta_0)$ does not satisfy the rank conditions, where $\theta_0 \in \Theta$ for global identifiability and $\theta_0 \in \Theta\setminus \bar{\Theta}$ for generic identifiability, where $\bar{\Theta} \subseteq \Theta$ is a set of measure zero. Denote $\mathbb{G}_\Theta= \{G(q,\theta)|\theta \in \Theta\}$. Then under Assumption~\ref{ass:openset}, there exists a positive real number $r$ such that $\mathcal{B}(G(q,\theta_0),r) \subseteq \mathbb{G}_\Theta$, where $\mathcal{B}(G(q,\theta_0),r)$ denotes the open ball in $\mathbb{G}^\star$ of center $G(q,\theta_0)$ and radius $r$. Then based on Lemma~\ref{lemma:nece1}, there exists a different network model $\bar{M}$ that corresponds to the same $T_{\mathcal{W}\mathcal{X}}$ as $M(\theta_0)$, and $\bar{G} \in \mathcal{B}(G(\theta_0),r) \subseteq \mathbb{G}_\Theta$. In addition, due to Assumption~\ref{ass:indePara}, there exists $\theta_1 \in \Theta$ such that $M(\theta_1) = \bar{M}$. This concludes that $T_{\mathcal{W}\mathcal{X}}(q,\theta_0)$ leads to both $G_{j\bar{\mathcal{W}}_j}(q,\theta_0)$ and $G_{j\bar{\mathcal{W}}_j}(q,\theta_1)$, which contradicts with global identifiability and generic identifiability.
$\hfill{} \blacksquare$
\end{pf}
%
\subsection{Proof for Proposition~\ref{prop:TsubRank}}

The following result is instrumental in the proof.

\begin{lemma} \label{prop:graph}
Consider a model set $\mathcal{M} $ that satisfies Assumptions~\ref{ass:indePara} and \ref{ass5}. Then for any subsets $\bar{\mathcal{W}} \subseteq \mathcal{W}$ and $\bar{\mathcal{X}} \subseteq \mathcal{X}$, $T_{\bar{\mathcal{W}}\bar{\mathcal{X}}}$  is full rank for almost all $\theta \in \Theta$ if and only if
$F(\bar{\mathcal{W}},\bar{\mathcal{X}})$, as defined in \eqref{eq:Fdfn}, is structural full rank.
\end{lemma}
The above result suggests that testing the generic rank of $T_{\bar{\mathcal{W}}\bar{\mathcal{X}}}$ only needs to evaluate the sparsity pattern of $F(\bar{\mathcal{W}},\bar{\mathcal{X}})$. Then due to the connection between graph $\mathcal{G}$ and $F$'s sparsity pattern, the above result allows us to obtain Proposition~\ref{prop:TsubRank}. To prove Lemma~\ref{prop:graph}, two preliminary results on structural rank are presented. Consider a block matrix whose entries are either zeros or distinct indeterminates:
\begin{equation}
M= \begin{bmatrix}
A-I & B \\
C & D
\end{bmatrix}, \label{eq:M}
\end{equation}
where $A$ is hollow, $A$ and $D$ are of dimensions $L \times L$ and $m \times m$, respectively. Here $L$ or $m$ is allowed to be zero.

\begin{lemma} \label{permuzero}
If $M$ in \eqref{eq:M} is \textbf{not} structural full rank, then $M$ can be permuted as
\begin{equation}
\begin{bmatrix}
\bar{A} & 0 \\
\bar{C} & \bar{D}
\end{bmatrix}, \label{lem:graph1}
\end{equation}
where $\bar{A}$ has dimension $k_1 \times (k_1 - 1)$ for some $k_1  \geqslant 1$.
\end{lemma}
\begin{pf}
The non-zero structure of $M$ can be characterized by a graph $\bar{\mathcal{G}}: = (\bar{\mathcal{V}}, \bar{\mathcal{E}})$ with $\bar{\mathcal{V}}: = \bar{\mathcal{W}} \cup  \bar{\mathcal{X}} \cup \bar{\mathcal{Y}}$, where $\mathcal{W} = \{w_1,\cdots,w_L\}$, $\bar{\mathcal{X}} = \{x_1,\cdots,x_m\}$, and $\mathcal{Y} = \{y_1,\cdots,y_m\}$ correspond to the rows/columns of $A$,  the columns of $B$ and  the rows of $C$, respectively. Besides, a directed edge $(j,i) \in \bar{\mathcal{E}}: = \bar{\mathcal{V}} \times \bar{\mathcal{V}}$ if and only if $M_{ij}$ is non-zero.
When $M$ is not structural full rank, it follows from \citep{van1991graph} that $b_{\bar{\mathcal{X}} \rightarrow \bar{\mathcal{W}}} < |\bar{\mathcal{X}}| = m$ in $\bar{\mathcal{G}}$.
Then, from Theorem~\ref{thm:Menger}, there exists a $\bar{\mathcal{X}} - \bar{\mathcal{W}}$ disconnecting set $\mathcal{D}$ with $|\mathcal{D}| = m-1$ in $\bar{\mathcal{G}}$. Note that with $\mathcal{D}$ and based on Lemma~\ref{lem0}, we can divide $\bar{\mathcal{V}}$ into three disjoint sets $\mathcal{D}$, $\mathcal{S}$ and $\mathcal{P}$ with $|\mathcal{S}|+|\mathcal{P}| = L+m+1$. Moreover, there is no edge from $\mathcal{S}$ to $\mathcal{P}$, where $\mathcal{S} \subseteq \bar{\mathcal{W}} \cup \bar{\mathcal{X}}$, $\mathcal{P} \subseteq \bar{\mathcal{W}} \cup \bar{\mathcal{Y}}$. Thus, from the definiton of $\bar{\mathcal{E}}$, we can find a permutation of $M$ in the form of \eqref{lem:graph1} with a zero block, whose rows and columns correspond to $\mathcal{P}$ and $\mathcal{S}$, respectively. Furthermore, the column dimension of $\bar{A}$ is computed as $L+m-|\mathcal{S}|= |\mathcal{P}|-1$, which completes the proof. $\hfill{} \blacksquare$
\end{pf}

\begin{lemma} \label{lemma:algefinal}
Let the non-zero entries of $M$ in \eqref{eq:M} be divided into two disjoint sets $\mathcal{P}_1$ and $\mathcal{P}_2$.
If $\det(M)$ does not depend on the entries in $\mathcal{P}_1$, then
$$
\det(M) = (-1)^j \prod_{i=1}^l \det(A_i), \ \text{for some}~l \geqslant 1, j \in \{0,1\},
$$
where $A_i$ is a square submatrix of $M$ that contains only non-zero entries in $\mathcal{P}_2$.
\end{lemma}
\begin{pf}
When $\mathcal{P}_1 = \emptyset$, the proof is trivial. Now suppose $\mathcal{P}_1 \ne \emptyset$, and consider the cofactor expansion formula of $\det(M)$, which contains a term $b \det(\bar{B})$
with $\det(\bar B)$ the cofactor of the nonzero entry $b \in \mathcal{P}_1$. Since $\det(M)$ does not depend on $b$, we obtain $\det(\bar{B}) =0$, i.e., $\bar{B}$ is not structural full rank. Then it follows from Lemma~\ref{permuzero} that
$M$ can be be permuted as
\begin{equation*}
    \begin{bmatrix}
    \star & \begin{bmatrix}
    \bar{B}_{11} & 0 \\ \bar{B}_{21} & \bar{B}_{22}
    \end{bmatrix} \\
    b & \star
    \end{bmatrix} : = \begin{bmatrix}
\tilde{A} & 0 \\
\tilde{C} & \tilde{D}
\end{bmatrix},
\end{equation*}
with square matrices $\tilde{A}$ and $\tilde{D}$, and $\tilde{C}$ containing $b$. Then, we have $\det(M) =  \pm \det(\tilde{A})\det(\tilde{D})$, which does not depend on $b$. Here, the sign $ \pm$ depends on the permutation. If $\tilde{A}$ or $\tilde{D}$ contains other entries in $\mathcal{P}_1$, the above analysis can be applied recursively, which proves the lemma. $\hfill{} \blacksquare$
\end{pf}
With the previous lemmas, Lemma~\ref{prop:graph} can be achieved.
\begin{pf}
Without loss of generality, consider $|\bar{\mathcal{W}}| \leqslant |\bar{\mathcal{X}}|$, and thus $T_{\bar{\mathcal{W}}\bar{\mathcal{X}}}$ having full generic rank implies that its square submatrix $T_{\bar{\mathcal{W}}\bar{\mathcal{X}}_1}$ has full rank generically. Consider $\bar{F}$ defined as \eqref{eq:F} for $T_{\bar{\mathcal{W}}\bar{\mathcal{X}}_1}$, and denote $|\bar{\mathcal{X}}_1|=m$. As $\bar{F}$ is structural full rank if and only if $F(\bar{\mathcal{W}},\bar{\mathcal{X}}_1)$ (and thus $F(\bar{\mathcal{W}},\bar{\mathcal{X}})$) has structural full rank, and $\bar{F}$ has generically full rank if and only if $T_{\bar{\mathcal{W}}\bar{\mathcal{X}}_1}$ has generically full rank. Thus, the proof aims to prove that $\bar{F}$ is structural full rank if and only if it is generically full rank. Let $a_{ij}$ denote the $(i,j)$ entry of $\bar{F}$. The ``if" part is clear as being structural full rank is a necessary condition for being full rank generically. For the ``only if" part, consider now all the summands in the Leibnitz formula for $\det(\bar{F})$. If $\bar{F}$ is structural full rank, without loss of generality, $\det(\bar{F})$ contains a term according to a permutations $\bar{\sigma}$ of $[1,\cdots,L+m]$ as
\begin{equation}
a\prod_{i=p+1}^{L+m} a_{\bar{\sigma}(i)i} \not= 0,\label{eq:final2}
\end{equation}
for some $p \geqslant 0$, where $\sigma(i)$ denotes the $i$-th index in the permutation, and $a$ is the product of all parametrized entries in the term and also contains the maximum number of parametrized entries among the other permutations of $[1,\cdots,L+m]$. In addition, the summation of all terms with the common factor $a$ in $\det(\bar{F})$ equals
\begin{equation}
a \det(\bar{F}_{\{\bar{\sigma}(p+1),\cdots,\bar{\sigma}(L+m)\} \{p+1,\cdots,L+m\}}). \label{eq:final3}
\end{equation}
Denote the above submatrix of $\bar{F}$ as $F_1$. Then $\bar{F}$ is proved to be generically full rank by showing that the term \eqref{eq:final3} is non-zero and cannot be canceled by the other terms in $\det(\bar{F})$ generically as follows. As \eqref{eq:final2} contains the maximum number of parametrized entries compared to the other permutations, $\det(F_1)$ does not dependent on the parametrized entries when considered as a polynomial of $F_1$'s all non-zero entries. Based on Lemma~\ref{lemma:algefinal} and the fact that $C$ is a selection matrix, \eqref{eq:final3} can be reformulated as
\begin{equation}
(-1)^j a \prod_{i=1}^l \det(A_i), \label{eq:proof222}
\end{equation}
for some $l \geqslant 1$ and $j \in \{0,1\}$, where $A_i$ contains only known transfer functions and is square submatrix of $F_1$ and thus of $F(\bar{\mathcal{W}},\bar{\mathcal{X}}_1)$ defined in \eqref{eq:Fdfn}. As \eqref{eq:proof222} contains the non-zero term \eqref{eq:final2} after expansion, $A_i$ is structural full rank and thus full rank by Assumption~5, for all $i$. This means that $\det(F_1)$ is non-zero and thus $\det(\bar{F})$ contains at least the non-zero term \eqref{eq:final3}. As each parametrized entry depends on independent parameters by Assumption~\ref{ass:indePara}, the non-zero term \eqref{eq:final3} cannot be canceled in $\det(\bar{F})$ by the other terms without the factor $a$. Thus, $\det(\bar{F})$ is a non-constant analytic function of $\theta$. Then following the property of analytic functions and the analysis in Lemma~V.2 of \citep{hendrickx2018identifiability}, $\bar{F}$ and thus $T_{\bar{\mathcal{W}}\bar{\mathcal{X}}}$ have full rank generically. $\hfill{} \blacksquare$
\end{pf}

Finally, Proposition~\ref{prop:TsubRank} can be proved based on Lemma~\ref{prop:graph}.
\begin{pf} The proof is analogous to the proof of Proposition~V.1 in \citep{hendrickx2018identifiability}: For generically $\rank(T_{\bar{\mathcal{W}}\bar{\mathcal{X}}}) \geqslant b_{\bar{\mathcal{X}} \to \bar{\mathcal{W}}}$, consider a subgraph of the network containing all the vertices but only the edges in a set of maximum number vertex disjoint paths from $\bar{\mathcal{X}}$ to $\bar{\mathcal{W}}$. Let $\bar{\mathcal{X}}_1 \subseteq \bar{\mathcal{X}}$ and $\bar{\mathcal{W}}_1 \subseteq \bar{\mathcal{W}}$ denote the starting and ending vertices of the vertex disjoint paths, respectively. The obtained subgraph's stucture can then be encoded by matrices $A$ and $B$ with only zeros and ones, where $A_{ji}=1$ and $B_{kn}=1$ if and only if $G_{ji}$ and $X_{kn}$ denote the edges in the subgraph, respectively.  Then following the same analysis of Proposition~V.1 in \citep{hendrickx2018identifiability}, we can show that $C(I-A)^{-1} B_{\mathcal{W} \bar{\mathcal{X}}_1} $ is a permutation matrix and thus has full rank that equals $b_{\bar{\mathcal{X}} \to \bar{\mathcal{W}}}$, where $C$ is a selection matrix that extracts rows of $(I-A)^{-1}$ corresponding to $\bar{\mathcal{X}}_1 $. Then following Lemma~\ref{lem:Fintro} similarly, it holds that
$$
\begin{bmatrix}
(A-I)_{\mathcal{W} (\mathcal{W} \setminus \bar{\mathcal{W}}_1)} &  B_{\mathcal{W} \bar{\mathcal{X}}_1}
\end{bmatrix}.
$$
is full rank, and thus $F(\bar{\mathcal{W}}_1,\bar{\mathcal{X}}_1)$ defined in \eqref{eq:Fdfn} is structural full rank. Then based on Lemma~\ref{prop:graph}, $T_{\bar{\mathcal{W}}_1\bar{\mathcal{X}}_1}$ is generically full rank that equals $b_{\bar{\mathcal{X}} \to \bar{\mathcal{W}}}$, and thus $\rank(T_{\bar{\mathcal{W}}\bar{\mathcal{X}}}) \geqslant b_{\bar{\mathcal{X}} \to \bar{\mathcal{W}}}$ generically.\\
For $rank(T_{\bar{\mathcal{W}}\bar{\mathcal{X}}}) \leqslant b_{\bar{\mathcal{X}} \to \bar{\mathcal{W}}}$ generically, a minimum $\bar{\mathcal{X}} - \bar{\mathcal{W}}$ disconnecting set can be considered as in \citep{hendrickx2018identifiability}, which leads to permuted network matrices with block zeros as also explored later in \eqref{matrix:sepe}. Then the proof can be achieved similarly as in \citep{hendrickx2018identifiability}. Note that \citep{hendrickx2018identifiability} requires its Lemma~V.2 to ensure the invertibility of certain submatrix $I-G_{\bar{\mathcal{W}}_1 \bar{\mathcal{W}}_1}$, which is guaranteed by Assumption~1(b) in this work and thus the lemma is not needed. $\hfill{} \blacksquare$
\end{pf}
%
%

\subsection{Proof of Theorem~\ref{theorem:SingleCutNew}}
Theorem~\ref{theorem:SingleCutNew} is established by two graphical results.
\begin{lemma} \label{lemma:VDpath0}
In a simple directed graph, given a set $\mathcal{P}$ of vertex disjoint paths from vertex set $\mathcal{V}_1$ to a vertex set $\mathcal{V}_2$, there exists a set $\mathcal{P}_{new}$ of vertex disjoint paths from $\mathcal{V}_1$ to $\mathcal{V}_2$ such that $|\mathcal{P}_{new}| = |\mathcal{P}|$ and paths in $\mathcal{P}_{new}$ are internally vertex disjoint\footnote{A path is internally vertex disjoint with a set of vertices $\mathcal{V}$, if the internal vertices of the path are not in $\mathcal{V}$.} with $\mathcal{V}_1 \cup \mathcal{V}_2 $.
\end{lemma}
\begin{pf}
We prove the lemma by showing that there always exists a $\mathcal{P}_{new}$ by modifying the paths in $\mathcal{P}$. Let $w_i \rightarrow w_j$ be an arbitrary path in $\mathcal{P}$ which contains an internal vertex in $\mathcal{V}$, then we can always replace $w_i \rightarrow w_j$ by its subpath which contains a starting vertex in $\mathcal{V}_1$ and an ending vertex in $\mathcal{V}_2$, while the other vertices in the subpath are not in $\mathcal{V}$. This includes the special case that the obtained subpath has no internal vertex. Applying the above modification to all the paths in $\mathcal{P}$, ,which contain internal vertices in $\mathcal{V}$, leads to $\mathcal{P}_{new}$. $\hfill{} \blacksquare$
\end{pf}

Based on the above result, there always exists a set of the maximum number of vertex disjoint paths from $\mathcal{V}_1$ and $\mathcal{V}_2$, which are internally vertex disjoint with $\mathcal{V}_1 \cup \mathcal{V}_2$. We denote such a set by $\mathcal{B}_{\mathcal{V}_1 \to \mathcal{V}_2}$. With this notation, an instrumental graphical result can be obtained.
\begin{lemma} \label{lemma:VdPath}
Consider a simple directed graph $\mathcal{G}=(\mathcal{V},\mathcal{E})$ with any two vertex sets $\mathcal{V}_1$, $\mathcal{V}_2$ $\subseteq \mathcal{V}$ and a subset $\bar{\mathcal{V}}_2 \subseteq \mathcal{V}_2$. The following statements are equivalent:
\begin{enumerate}
\item $b_{\mathcal{V}_1 \to \bar{\mathcal{V}}_2 } = | \bar{\mathcal{V}}_2 | $ and $b_{\mathcal{V}_1 \to \mathcal{V}_2 } = b_{\mathcal{V}_1 \to \bar{\mathcal{V}}_2 } + b_{\mathcal{V}_1\to \mathcal{V}_2 \setminus \bar{\mathcal{V}}_2} $;

\item there exists a $\mathcal{V}_1  - \mathcal{V}_2 \setminus \bar{\mathcal{V}}_2$ disconnecting set $\mathcal{D}$ such that
\begin{equation}
b_{\mathcal{V}_1 \to \mathcal{D} \cup \bar{\mathcal{V}}_2} = |\mathcal{D}| + |\bar{\mathcal{V}}_2|; \label{eq:disconnecLemma1}
\end{equation}

\item there exist a $\bar{\mathcal{V}}_1 \subseteq \mathcal{V}_1$ and a $\bar{\mathcal{V}}_1 - \mathcal{V}_2 \setminus \bar{\mathcal{V}}_2$ disconnecting set $\mathcal{D}$ such that
$
b_{\bar{\mathcal{V}}_1 \to \bar{\mathcal{V}}_2 \cup \mathcal{D}} = |\mathcal{D}| + |\bar{\mathcal{V}}_2|.
$
\end{enumerate}
\end{lemma}
\begin{pf}
We first prove that (1) holds if and only if (2) holds. If (2) holds, \eqref{eq:disconnecLemma1} shows that there exist $|\mathcal{D}|$ and $|\bar{\mathcal{V}}_2|$ vertex disjoint paths from $\mathcal{V}_1$ to $\mathcal{D}$ and from $\mathcal{V}_1$ to $\bar{\mathcal{V}}_2$, respectively. Then according to Lemma~\ref{lemma:VDpath0}, there exist $\mathcal{B}_{\mathcal{V}_1  \to \bar{\mathcal{V}}_2}$ with cardinality $|\bar{\mathcal{V}}_2|$ and $\mathcal{B}_{\mathcal{V}_1  \to \mathcal{D} }$ with cardinality of $|\mathcal{D}|$, and the paths in the above two sets are vertex disjoint. A new set $\mathcal{B}_{\mathcal{D}  \to \mathcal{V}_2 \setminus \bar{\mathcal{V}}_2 }$ can also be introduced and is vertex disjoint with both $\mathcal{B}_{\mathcal{V}_1  \to \bar{\mathcal{V}}_2}$ and $\mathcal{B}_{\mathcal{V}_1 \to \mathcal{D} }$, because if not vertex disjoint, a path from $\mathcal{V}_1$ to $\mathcal{V}_2 \setminus \bar{\mathcal{V}}_2 $ will exist and do not intersect with $\mathcal{D}$, contradicting $\mathcal{D}$ as a disconnecting set. Then by linking a subset of paths in $\mathcal{B}_{\mathcal{V}_1 \to \mathcal{D} }$ and all paths in $\mathcal{B}_{\mathcal{D}  \to \mathcal{V}_2 \setminus \bar{\mathcal{V}}_2}$, we can obtain a set of vertex disjoint paths from $\mathcal{V}_1$ to $\mathcal{V}_2 \setminus \bar{\mathcal{V}}_2$. Since $\mathcal{D}$ is a $\mathcal{V}_1 - \mathcal{V}_2 \setminus \bar{\mathcal{V}}_2$ disconnecting set and $|\mathcal{B}_{\mathcal{V}_1 \to \mathcal{D}}| =|\mathcal{D}|$, the obtained set of paths forms a set $\mathcal{B}_{\mathcal{V}_1 \to \mathcal{V}_2 \setminus \bar{\mathcal{V}}_2}$ with the maximum number of vertex disjoint paths, and its cardinality equals $|\mathcal{B}_{\mathcal{D}  \to \mathcal{V}_2 \setminus \bar{\mathcal{V}}_2}|$. In addition, the obtained $\mathcal{B}_{\mathcal{V}_1 \to \mathcal{V}_2 \setminus \bar{\mathcal{V}}_2}$ is also vertex disjoint with $\mathcal{B}_{\mathcal{V}_1  \to \bar{\mathcal{V}}_2}$. Since $b_{\mathcal{V}_1 \to \mathcal{V}_2 } \leqslant b_{\mathcal{V}_1 \to \bar{\mathcal{V}}_2 } + b_{\mathcal{V}_1\to \mathcal{V}_2 \setminus \bar{\mathcal{V}}_2} $ always holds, $\mathcal{B}_{\mathcal{V}_1  \to \bar{\mathcal{V}}_2}$ with cardinality $|\bar{\mathcal{V}}_2|$ and $\mathcal{B}_{\mathcal{V}_1 \to \mathcal{V}_2 \setminus \bar{\mathcal{V}}_2}$ together forms a set of maxinum number of vertex disjoint paths from $\mathcal{V}_1$ to $\mathcal{V}_2$, which proves (1).\\
If (1) holds, let $\mathcal{D}$ be a minimum $\mathcal{V}_1-  \mathcal{V}_2 \setminus \bar{\mathcal{V}}_2$ disconnecting set, and we have $\mathcal{B}_{\mathcal{V}_1  \to \bar{\mathcal{V}}_2}$ with cardinality $|\bar{\mathcal{V}}_2|$ and $\mathcal{B}_{\mathcal{V}_1  \to \mathcal{D}}$ with cardinality $|\mathcal{D}|$ which are vertex disjoint. Thus the above two sets together form a set of vertex disjoint paths from $\mathcal{V}_1$ to $\bar{\mathcal{V}}_2 \cup \mathcal{D}$, which leads to
$
b_{\mathcal{V}_1  \to \bar{\mathcal{V}}_2 \cup \mathcal{D}} \geqslant |\mathcal{D}| + |\bar{\mathcal{V}}_2|.
$
As $b_{\mathcal{V}_1  \to \bar{\mathcal{V}}_2 \cup \mathcal{D}}$ is upper bounded by $|\mathcal{D}| + |\bar{\mathcal{V}}_2|$, it then holds that $b_{\mathcal{V}_1  \to \bar{\mathcal{V}}_2 \cup \mathcal{D}} = |\mathcal{D}| + |\bar{\mathcal{V}}_2|$, which leads to $(1) \iff	(2)$.
\\
Then we shown that (2) is equivalent to (3). The implication $(2) \implies (3)$ is straightforward by letting $\bar{\mathcal{V}}_1 = \mathcal{V}_1$. For $(3) \implies (2)$,  if $(3)$ holds, then $\mathcal{D}_1 =  \mathcal{D} \cup (\mathcal{V}_1 \setminus \bar{\mathcal{V}}_1 )$ becomes a $\mathcal{V}_1 - \mathcal{V}_2 \setminus \bar{\mathcal{V}}_2$ disconnecting set, and $b_{\mathcal{V}_1  \to \mathcal{D}_1  \cup \bar{\mathcal{V}}_2} = |\mathcal{D}_1| + |\bar{\mathcal{V}}_2|$ since a single vertex can be regarded to have a path to itself, which concludes the proof. $\hfill{} \blacksquare$
\end{pf}
Finally, Lemma~\ref{lemma:VdPath} and Theorem~\ref{theorem:PathCondi} lead to Theorem~\ref{theorem:SingleCutNew}.

\subsection{Proof of Proposition~\ref{propo:parallel}}
Before proving the proposition, we first prove that there exists a directed path from $w_i$ to $\mathcal{W}_j \setminus \{w_i\}$ if and only if there exist a parallel path from $w_i$ to $w_j$ or a cycle around the output $w_j$.  Note that due to model (\ref{eq:model}), $\mathcal{G}$ is a \textit{simple graph}, i.e. there is no self-loop such as $(w_i,w_i)$, and no parallel directed edges from one vertex to another vertex. For ``if" part, if there exists a parallel path from $w_i$ to $w_j$, this parallel path has to intersect with $\mathcal{W}_j  \setminus \{w_i\}$. Then we can find a directed path from $w_i$ to one vertex in $\mathcal{W}_j \setminus \{w_i\}$ as a subpath of the parallel path. If a cycle around $w_j$ exists, it will also intersect with $\mathcal{W}_j  \setminus \{w_i\}$, and thus the cycle contains a subpath from $w_j$ to one vertex in $\mathcal{W}_j  \setminus \{w_i\}$. Linking this subpath and the edge $(w_i,w_j)$ leads to a path from $w_i$ to $\mathcal{W}_j  \setminus \{w_i\}$.\\
For ``only if" part, for any directed path from $w_i$ to $w_k \in \mathcal{W}_j  \setminus \{w_i\}$, if the path does not contain edge $(w_i,w_j)$, then combining the the path and the edge $(w_k,w_j)$ will create a parallel path. If the path contains $(w_i,w_j)$, then combining the path and the edge $(w_k,w_j)$ while excluding $(w_i,w_j)$ will lead to a cycle around $w_j$. This concludes the relationship between the parallel paths, the cycles around the output and the paths from $w_i$ to $\mathcal{W}_j \setminus \{w_i\}$.

Then based on the above result, the ``only if" of the proposition is straightforward. For the ``if" part, if we collect an internal vertex from each parallel path and a vertex from cycle around the output into $\mathcal{D}$, $\mathcal{D}$ then must disconnect from $w_i$ to $\mathcal{W}_j  \setminus \{w_i\}$.

\subsection{Proof of Theorem~\ref{the:Decompo2}}
According to Lemma~\ref{lem0}, the disconnecting set separates all the vertices in the graph into three disjoint sets as $\mathcal{V} = \mathcal{S} \cup \mathcal{D} \cup \mathcal{P}$, while there is no directed edge from $\mathcal{S}$ to $\mathcal{P}$. In addition, each set may contain both internal signals and external signals, i.e. $\mathcal{S} = \mathcal{S}_x \cup \mathcal{S}_w$, $\mathcal{D} = \mathcal{D}_x \cup \mathcal{D}_w$ and $\mathcal{P} = \mathcal{P}_x \cup \mathcal{P}_w$, and it holds that $\bar{\mathcal{X}} \subseteq \mathcal{S}_x \cup \mathcal{D}_x$ and $\bar{\mathcal{W}} \subseteq \mathcal{D}_w \cup \mathcal{P}_w$. Algebraically, the above statements mean that there exist permuted network matrices such that
\begin{align}
 G &= \begin{bmatrix}
G_{\mathcal{S}_w \mathcal{S}_w} & G_{\mathcal{S}_w \mathcal{D}_w} & G_{\mathcal{S}_w \mathcal{P}_w} \\
G_{ \mathcal{D}_w \mathcal{S}_w} & G_{\mathcal{D}_w \mathcal{D}_w} & G_{\mathcal{D}_w \mathcal{P}_w} \\
0 & G_{\mathcal{P}_w \mathcal{D}_w} & G_{\mathcal{P}_w \mathcal{P}_w}
\end{bmatrix},\nonumber \\ X &= \begin{bmatrix}
X_{\mathcal{S}_w \mathcal{S}_x} & X_{\mathcal{S}_w \mathcal{D}_x} & X_{\mathcal{S}_w \mathcal{P}_x} \\
X_{ \mathcal{D}_w \mathcal{S}_x} & X_{\mathcal{D}_w \mathcal{D}_x} & X_{\mathcal{D}_w \mathcal{P}_x}  \\
0 & X_{\mathcal{P}_w \mathcal{D}_x} & X_{\mathcal{P}_w \mathcal{P}_x}
\end{bmatrix}, \label{matrix:sepe}
\end{align}
and $T_{\mathcal{W}\mathcal{X}}$ can be partitioned similarly without zero block matrices, where, for example, $T_{\mathcal{S}_w \mathcal{D}_x}$ denotes the transfer matrix from $\mathcal{D}_x$ to $\mathcal{S}_w$. Based on the above structure and the equation $(I-G)T_{\mathcal{W}\mathcal{X}}=X$, our goal is to find a proper matrix $K$ such that $T_{\bar{\mathcal{W}} \bar{\mathcal{X}}}= K T_{\mathcal{D}\bar{\mathcal{X}}}$. \\
Firstly, considering the division of the sets $\mathcal{D}$ and $\bar{\mathcal{X}}$ ,the mapping $T_{\mathcal{D}\bar{\mathcal{X}}}$ can be re-written as
\begin{equation}
T_{\mathcal{D}\bar{\mathcal{X}}} = \begin{bmatrix}
T_{\mathcal{D}_w \bar{\mathcal{X}}_S} & T_{\mathcal{D}_w \bar{\mathcal{X}}_D} \\
0 & I \\
0 & 0
\end{bmatrix}, \label{eq:appen1}
\end{equation}
where $\bar{\mathcal{X}}_S = \bar{\mathcal{X}} \cap \mathcal{S}_x$ and $\bar{\mathcal{X}}_D = \bar{\mathcal{W}} \cap \mathcal{D}_x$, the identity matrix is the mapping $T_{\bar{\mathcal{X}}_D \bar{\mathcal{X}}_D}$. Note that the rows of the bottom block matrices in \eqref{eq:appen1} correspond to the vertices in $\mathcal{D}_x \setminus \bar{\mathcal{X}}_D$. In addition, $T_{\bar{\mathcal{W}} \bar{\mathcal{X}}}$ can be written as
\begin{equation}
T_{\bar{\mathcal{W}} \bar{\mathcal{X}}}= \begin{bmatrix}
T_{\bar{\mathcal{W}}_P \bar{\mathcal{X}}} \\
T_{\bar{\mathcal{W}}_D \bar{\mathcal{X}}}
\end{bmatrix}, \label{eq:appen2}
\end{equation}
where $\bar{\mathcal{W}}_P = \bar{\mathcal{W}} \cap \mathcal{P}_w$ and $\bar{\mathcal{W}}_D = \bar{\mathcal{W}} \cap \mathcal{D}_w$. Thus, it is clear that
\begin{equation}
T_{\bar{\mathcal{W}}_D \bar{\mathcal{X}}} = C \begin{bmatrix}
T_{\mathcal{D}_w \bar{\mathcal{X}}_S} & T_{\mathcal{D}_w \bar{\mathcal{X}}_D}
\end{bmatrix},  \label{eq:appen3}
\end{equation}
where $C$ is a selection matrix that extracts the rows of $\begin{bmatrix}
T_{\mathcal{D}_w \bar{\mathcal{X}}_S} & T_{\mathcal{D}_w \bar{\mathcal{X}}_D}
\end{bmatrix}$ corresponding to $\bar{\mathcal{W}}_D$. In addition, from the permuted matrices and the equation $(I-G)T_{\mathcal{W}\mathcal{X}}=X$, it holds that
$
T_{\mathcal{P}_w \mathcal{S}_x} = (I-G_{\mathcal{P}_w\mathcal{P}_w})^{-1}G_{\mathcal{P}_w \mathcal{D}_w}T_{\mathcal{D}_w \mathcal{S}_x},
$
where $I-G_{\mathcal{P}_w \mathcal{P}_w}(z)$ is invertible and inversely proper because $I-G_{\mathcal{P}_w \mathcal{P}_w}(z)$ is proper and the network is well-posed, i.e. $\lim_{z \to \infty} \det(I-G_{\mathcal{P}_w \mathcal{P}_w}(z)) \not= 0$. The above equation leads to
\begin{equation}
T_{\bar{\mathcal{W}}_P \bar{\mathcal{X}}_S} = \bar{K} T_{\mathcal{D}_w \bar{\mathcal{X}}_S}, \label{eq:appen3a}
\end{equation}
where $\bar{K}= [(I-G_{\mathcal{P}_w\mathcal{P}_w})^{-1}]_{\bar{\mathcal{W}}_P \star}G_{\mathcal{P}_w \mathcal{D}_w}$. Then combining the above equation with \eqref{eq:appen1}, \eqref{eq:appen2} and \eqref{eq:appen3} leads to
\begin{align*}
T_{\bar{\mathcal{W}} \bar{\mathcal{X}}} &=\begin{bmatrix}
 T_{\bar{\mathcal{W}}_P \bar{\mathcal{X}}_S} & T_{\bar{\mathcal{W}}_P \bar{\mathcal{X}}_D} \\
 T_{\bar{\mathcal{W}}_D \bar{\mathcal{X}}_S} & T_{\bar{\mathcal{W}}_D \bar{\mathcal{X}}_D}
\end{bmatrix} = \begin{bmatrix}
\bar{K} & T_{\bar{\mathcal{W}}_P \bar{\mathcal{X}}_D} -\bar{K}T_{\mathcal{D}_w \bar{\mathcal{X}}_D} & 0\\
C& 0 & 0
\end{bmatrix} \\
& \times \begin{bmatrix}
T_{\mathcal{D}_w \bar{\mathcal{X}}_S} & T_{\mathcal{D}_w \bar{\mathcal{X}}_D} \\
0 & I \\
0 & 0
\end{bmatrix} = KT_{\mathcal{D}\bar{\mathcal{X}}}.
\end{align*}
The formulation of the above $K$ matrix can be further simplified. Based on the permuted matrices and the equation $(I-G)T_{\mathcal{W}\mathcal{X}}=X$, it holds
$
(I-G_{\mathcal{P}_w \mathcal{P}_w})T_{\mathcal{P}_w \mathcal{D}_x} - G_{\mathcal{P}_w \mathcal{D}_w} T_{\mathcal{D}_w \mathcal{D}_x}=X_{\mathcal{P}_w \mathcal{D}_x}.
$
Thus, we can conclude that
\begin{equation}
K = \begin{bmatrix}
\bar{K} & [(I-G_{\mathcal{P}_w \mathcal{P}_w})^{-1}]_{\bar{\mathcal{W}}_P \star} X_{\mathcal{P}_w \bar{\mathcal{X}}_D}& 0\\
C& 0 & 0
\end{bmatrix}, \label{eq:finalK}
\end{equation}
where $\bar{K}$ is defined in \eqref{eq:appen3a}; $C$ is defined in \eqref{eq:appen3} and its rows correspond to $\bar{\mathcal{W}}_D$; the columns of the last block column in \eqref{eq:finalK} correspond to $\mathcal{D}_x \setminus \bar{\mathcal{X}}_D$. Note that certain blocks in $K$ may disappear depending on if the corresponding set of signals is empty. $\hfill{} \blacksquare$

\bibliographystyle{plain}
\bibliography{allocation}             

\end{document}